\documentclass[10pt,twocolumn, notitlepage, superscriptaddress,prb, longbibliography,floatfix]{revtex4-2}

\usepackage{siunitx,booktabs}
\usepackage{tabularx}
\usepackage[normalem]{ulem}
\usepackage[utf8]{inputenc}
\usepackage[export]{adjustbox}
\usepackage{braket}
\usepackage{float}
\usepackage{amssymb}
\usepackage{makecell}
\usepackage[intlimits]{amsmath}
\usepackage{graphicx}
\usepackage{here}
\usepackage{csquotes}

\usepackage{tabularx}
\usepackage{tikz}
\usepackage{siunitx,booktabs}
\usepackage{makecell}
\usepackage{array,varwidth,lipsum}
\usepackage{booktabs}
\usepackage{siunitx}
\usepackage{color}
\usepackage[toc,page]{appendix}
\usepackage[colorlinks=true,linkcolor=Blue,citecolor=Blue,filecolor=Blue]{hyperref}

\newcolumntype{d}[1]{D{.}{.}{#1}}
\newcommand\mc[1]{\multicolumn{1}{c}{#1}}
\hypersetup{
    colorlinks=true,
    urlcolor=blue,
    citecolor=blue,
    linkcolor=blue}

\newcommand{\thickcline}[1]{%
    \@thickcline #1\@nil%
}
\newcommand{\ev}[1]{\left\langle{#1}\right\rangle}

\usepackage{etoolbox}           % <--
\renewcommand{\bfseries}{\fontseries{b}\selectfont} % <--
\robustify\bfseries             % <--
\newrobustcmd{\B}{\bfseries}

\newcommand{\thickhline}{%
    \noalign {\ifnum 0=`}\fi \hrule height 0.0pt
    \futurelet \reserved@a \@xhline
}
\newcolumntype{"}{@{\hskip\tabcolsep\vrule width 1.0pt\hskip\tabcolsep}}

\providecommand{\keywords}[1]{\textbf{\textit{Keywords}} #1}
\usepackage{booktabs}
\newcolumntype{C}[1]{>{\centering\let\newline\\\arraybackslash\hspace{0pt}}m{#1}}

\begin{document}

\title{Investigation of magnetic properties of $4f$-adatoms on graphene}

\author{Johanna P. Carbone}
\email{j.carbone@fz-juelich.de}
\affiliation{Peter Gr\"unberg Institut and Institute for Advanced Simulation, Forschungszentrum J\"ulich and JARA, 52425 J\"ulich, Germany \looseness=-1}
\affiliation{Physics Department, RWTH-Aachen University, 52062 Aachen, Germany}
\author{Juba Bouaziz}
\affiliation{Peter Gr\"unberg Institut and Institute for Advanced Simulation, Forschungszentrum J\"ulich and JARA, 52425 J\"ulich, Germany \looseness=-1}
\author{Gustav Bihlmayer}
\affiliation{Peter Gr\"unberg Institut and Institute for Advanced Simulation, Forschungszentrum J\"ulich and JARA, 52425 J\"ulich, Germany \looseness=-1}

\author{Stefan Bl\"{u}gel}
\affiliation{Peter Gr\"unberg Institut and Institute for Advanced Simulation, Forschungszentrum J\"ulich and JARA, 52425 J\"ulich, Germany \looseness=-1}

\begin{abstract}
Rare-earth (RE) atoms on top of 2D materials represent an interesting platform 
with the prospect of tailoring the magnetic anisotropy for practical applications. 
Here, we investigate the ground state and magnetic properties of selected $4f$-atoms 
deposited on a graphene substrate in the framework of the DFT+$U$ approach. The inherent 
strong spin-orbit interaction in conjunction with crystal field effects acting on the 
localized $4f$-shells results in a substantial magnetic anisotropy energy (tens of meVs), 
whose angular dependence is dictated by the $C_{6v}$ symmetry of the graphene substrate. 
We obtain the crystal field parameters and investigate spin-flip events via quantum 
tunneling of magnetization in the view of achieving a protected quantum-spin behavior. Remarkably, the large spin and orbital moments of the open $4f$-shells (Dy, Ho and Tm) generate 
a strong magneto-elastic coupling which provides more flexibility  to control the magnetic 
state via the application of external strain.

\end{abstract}
\keywords{rare-earths, graphene, magnetic anisotropy, first-principles calculations, crystal field effects}

\maketitle

\maketitle

\section{Introduction}

Graphene (Gr) is the first 2D-material to be discovered and stimulated interest at
the technological and fundamental research levels. A multitude of fascinating 
electronic phenomena can be tailored in different 2D-materials by tuning the 
chemical composition and structural properties~\cite{guo2021stacking,Miro2014,manzeli20172d,novoselov2016,liu2016van,geim2013van}. 
The experimental realization of these novel nanostructured systems might lead 
to the next-generation of efficient spintronics  devices~\cite{ahn20202d}.
In this material class, magnetic rare-earth (RE) atoms deposited on surfaces 
represent a promising pathway to achieve magnetic remanence, hence design 
stable memory devices. The rare-earth  localized $4f$ electrons carry
large spin and orbital moments, generating strong spin-orbit coupling effects (SOC)~\cite{jensen1991rare} and magneto-crystalline anisotropy, which, combined with 
a relatively low magnetic damping~\cite{schuh2012magnetic} results in stable nanometric 
scale magnets. More exotic magnetic phenomena can be generated 
and controlled when these rare-earth atoms are deposited on 2D-materials.

Also, at the experimental level, single-ion magnets deposited on top of  
2D-material substrates are under scrutiny~\cite{Gambardella,baltic2016superlattice,schuh2012magnetic,miyamachi2013stabilizing}. 
These might serve as building blocks for quantum computer "qubits", where the central
desired property is a long coherence time, which translates into a large magnetic
anisotropy in conjunction with a low damping of the magnetic excitations, thus
reducing the quantum fluctuation~\cite{ibanez2016zero,Bouaziz2020}. This raises
the prospect of employing the $4f$-elements as a quantistic information carrier~\cite{PRXQuantum.2.010312,bertaina2007rare}, since the magnetic unit 
emerges from strongly localized $4f$-electrons well-separated from the surrounding
itinerant $spd$-electron cloud. The chemical and magnetic interactions of RE 
atoms on 2D-materials have also received attention~\cite{3d4fHuttman,Pivettameasure,Liu2010,Tm/W,DySTO,4focc}, 
and how the protection of RE atoms against quantum tunneling of magnetization~\cite{Baltic2018,BruneHoEr,Rusponithickness} can be exploited.

From the perspective of applications, tailoring the magnetic anisotropy is a
fundamental aspect. {For instance, Herman \textit{et al.}~\cite{herman2022tailoring}, 
investigated the possibility of modifying the magnetic anisotropy of a Dy/Ir surface alloy
by a lifting skyhook effect of the Dy 
atoms from the Ir surface when brought in contact with a graphene sheet. Here, the $fd$ hybridization of the Dy electrons generates long tails of the charge distribution that overlap with the Ir(111) orbitals. When graphene is deposited on-top of Dy/Ir(111) the $fd$ states of Dy hybridize also with the $\pi$ orbitals arising from the carbon structure, and this interaction results in an increased separation of the Dy atoms from the Ir(111) surface due to the Dy-Gr interaction. This effect leads to a redistribution of the charge density that enhances the magnetic anisotropy.}

Concerning single rare-earth atoms on surfaces, a graphene sheet is often introduced as a decoupling layer between a metallic 
or insulating surface and the RE atoms, in order to decouple the magnetic source 
from any possible scattering events for example of conduction electrons or phonons.
In addition, graphene presents a C$_{6v}$ symmetry which can be exploited to 
tune the quantum states of the RE atom in order to achieve further stabilization of the 
magnetization, since the respective hexagonal crystal field removes the degeneracy of the magnetic states in a free atom and thus can generate an energy dispersion that is protected against magnetization reversal. This magnetic stabilization strictly depends on the properties of the chosen RE atom, \textit{i.e.}\ its orbital and spin angular momentum, which are coupled by the spin-orbit coupling that determines the number and nature of magnetic states interacting with the crystal field.

Recently, several theoretical investigations have been carried out. From this perspective, the presence of strongly localized $4f$-orbitals enhances the complexity of first
principle approaches, since common approximations to describe the exchange and correlation energy such as the local density 
approximations (LDA) fail to provide an accurate description. To properly
account for the strong Coulomb effects in the $4f$-atoms, methods going beyond
the standard approximations to density functional theory (DFT), including strongly
correlated electron methods such as the Hubbard-I 
approximation~\cite{Peters2014,ShickHoPt(111),SHICK2019,shick2009HIA} or dynamical 
mean-field theory (DMFT)~\cite{held2007electronic}, are often adopted. A simpler alternative for the treatment of $4f$-electrons is the LDA+$U$ approach~\cite{Anisimov:1997}, 
which incorporates the local Coulomb repulsion in the form of a Hubbard correction 
in addition to the LDA exchange-correlation.

In the present work, we perform an analysis of magnetic properties of three 
heavy RE atoms namely Dy, Ho and Tm. The choice of these candidates is motivated by preceding experimental investigations~\cite{Baltic2018,Gambardella,Pivettameasure,Tm/W,baltic2016superlattice,Pivetta2018}. 
Using a supercell approach, we determine the electronic structure of these $4f$-atoms deposited on a free-standing graphene sheet. We employ the GGA+$U$ method 
parameterized following the formulation of Ref.~\cite{Shick1999}. The magnetic 
anisotropy constants are determined by fitting the total energy for different 
magnetization orientations relative to the crystal lattice. We then propose an analytical method which permits to 
reverse-engineer the crystal field parameters (CFP)  from the anisotropy 
constants. 
The calculated CFP are employed to construct the $C_{6v}$ crystal field matrix for each RE atom, which is diagonalized to obtain the $J_z$ multiplet spectrum of the RE/Gr complexes. We then examine the magneto-elastic coupling in terms of magnetic anisotropy constants by simulating an external stress acting on the samples with magnetizations aligned along different orientations and determine the frequency of the respective vibrational mode. Finally, we underline the importance of an accurate theoretical description of the $4f$-electrons by analyzing deviations of the magnetic anisotropy following different orbital occupations in the $4f$-shell.

 \section{Structural and electronic properties}
\label{parameters}

\subsection{Computational Details}
\begin{figure}[t]
    \centering
    \includegraphics[scale=0.15]{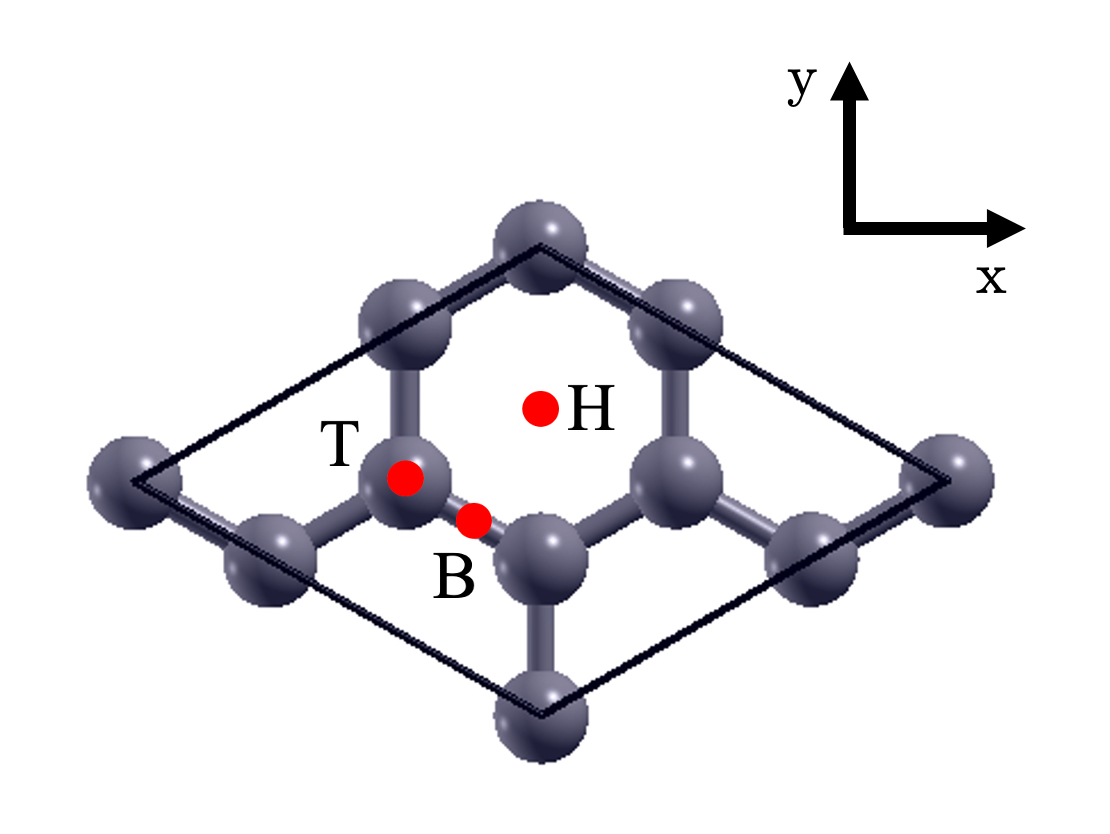}
    \caption{$\sqrt{3}\times\sqrt{3}$ supercell including $6$ Carbon 
    atoms (gray spheres). The three possible adsorption sites of the 
    rare-earth atom on graphene are depicted with red circles named \textquote{H}, \textquote{T} and \textquote{B}, respectively.}
    \label{Gr-sites}
\end{figure}

The presented results are obtained using DFT as implemented within 
the FLAPW (Full Potential Augmented Plane Wave) method using the FLEUR 
code~\cite{FLEUR,wortmann2023fleur}. All the simulations have been performed in a $\sqrt{3}\times\sqrt{3}$ supercell containing one magnetic atom and 
$6$ Carbon (C) atoms (see Fig.~\ref{Gr-sites}), with lattice constant equal to the experimental value of $2.46$ \AA, respectively multiplied by $\sqrt{3}$. The selection of the simulation cell is guided by both experimental and theoretical findings, as documented in Ref.~\cite{jugovac2023inducing,forster2012phase}. These suggest that Eu atoms, which share chemical characteristics with other rare-earth elements lacking an external $5d$ electron in the valence shell, tend to form a $\sqrt{3}\times\sqrt{3}$ superstructure on graphene. The spin-orbit coupling 
(SOC) was incorporated self-consistently adopting the second variation~\cite{2nd_variation} formulation on a $20\times 20$ 
$\mathbf{k}$-point mesh and a cut-off for the plane-wave basis functions
of $K_\mathrm{max}=4.5\, a_0^{-1}$ and $G_\mathrm{max}=13.5\, a_0^{-1}$ 
($a_0$ being the Bohr radius). The muffin-tin radii have been set to $2.80\, a_0$ for the RE atoms, and
$1.27\, a_0$  for the C atoms. The upper limit of the angular momentum 
inside the muffin-tin is set to $l_\mathrm{max}=10$ for the RE atom and 
$l_\mathrm{max}=6$ for C.
The exchange-correlation potential is taken 
in the generalized gradient approximation (GGA) following the parametrization 
PBE~\cite{PBE}. For the RE's correlated $4f$-orbitals, a Hubbard correction is applied, both Coulomb $U$ and intratomic exchange interaction $J$ are
included, the double counting is taken in fully localized limit (FLL)~\cite{DFTUShick}. 

\begin{table}[t]
\centering
\caption{\label{table1}  Adsorption energies (eV), for the different rare-earth atoms on graphene for each considered adsorption site. The calculations have been performed including SOC self-consistently following the formulation in equation ~\ref{Eads}.}

\begin{tabular*}{\linewidth}{c@{\extracolsep{\fill}}ccc}%{@{}SSSS}
\midrule[0.4mm]
$Site$& Dy& Ho &Tm\\
\midrule[0.2mm]
H& $-0.545$ &$-0.476$& $-0.399$\\
T& $-0.074$ &$-0.339$& $-0.280$\\
B& $-0.086$ &$-0.344$&$-0.286$\\
\midrule[0.4mm]
\end{tabular*}
%\end{ruledtabular}
%\end{indented}
\end{table}

We consider three heavy rare-earth prototypes namely: Dy, Ho and Tm. 
The values of the DFT+$U$ parameters are set to $U=7.0, \ 7.03,  \ 7.1$ eV and
$J=0.87, \ 0.83, \ 0.86$, respectively. These parameters are chosen following Ref.~\cite{ShickHoPt(111)} for Ho, and Ref.~\cite{SHICK2019} for Dy. The values Tm are chosen according to the interpolation formula given in Ref.~\cite{UandJ}. Moreover, the values of $U\sim 7$ eV are also chosen on the basis of the work conducted in ~\cite{Locht2018}, where it is shown that it is able to reproduce accurately the electronic and cohesive properties of RE bulk systems. We study a situation in which the electrons of the magnetic atoms relax into $4f$-occupations that reflect Hund's rules, \textit{i.e.}\ total angular momentum $J=8$ for Dy, $J=15/2$ for Ho and $J=7/2$ for Tm. We note that a deeper analysis shows that for Dy/Gr a solution deviating from Hund's rules with a total angular momentum of $J=8$ is found to be energetically more favorable. We present a detailed analysis of this state in Section~\ref{Dy5}.

\subsection{Structural Details}

In order to identify the lowest energy adsorption site, three possible
positions are taken into account: the hollow site (\textquote{H}) at the center of the hexagonal 
ring, on top of a C atom (\textquote{T}) and in the middle of a C-C bonding (\textquote{B}). These 
positions are illustrated in Fig.~\ref{Gr-sites} and the respective adsorption 
energies (in eV) are summarized in Table~\ref{table1}. The adsorption energies 
have been obtained considering the total energy difference between the total interacting 
system and the individual components as
\begin{equation}
    E_{ads}=E_{\text{RE/Gr}}-E_{\text{RE}}-E_{\text{Gr}}
    \label{Eads}
\end{equation}
in order to capture the energy involved in the formation of the complex, compared to the energy of the sum of the isolated RE atom and graphene monolayer. 
In Eq.~\ref{Eads} $E_\mathrm{RE/Gr}$ is the total energy of the RE/Gr complex, while $E_\mathrm{RE}$ and $E_\mathrm{Gr}$ correspond to the total energy of the isolated RE atom and graphene monolayer, respectively. 
The results show that for all three RE atoms
the H-site is energetically the most favorable, in agreement with several other theoretical and experimental studies~\cite{Liu2012,Liu2010,Baltic2018, baltic2016superlattice,Pivettameasure, Pivetta2018}, and the adatoms on graphene are described by the point group $C_{6v}$. The bonding strengths appear to be 
reduced on the T and B sites for Ho and Tm, and negligible for Dy.

%Experimental studies \cite{Pivettameasure, baltic2016superlattice, Pivetta2018} on rare-earth atoms adsorption behavior on graphene placed on an Ir(111) substrate reveal distinct temperature-dependent scenarios. Lower temperatures result in a disordered state due to inhibited adatom diffusion. At a critical temperature specific to each rare-earth element, surface diffusion toward favorable Moiré sites is activated, resulting in the formation of ordered superlattices. As the temperature continues to increase,
%the degree of order diminishes again.
%Additionally, these studies have quantified the energy barrier for diffusion, particularly the transition from one H-site to an adjacent lattice site, estimating it to be approximately $\sim 75$ meV for Dy.
%In parallel, our DFT calculations have examined total energy differences among various adsorption sites, including the H-site, B-site, and T-site. These computed values, approximately $\sim 0.1$ eV, are consistent with Liu et al.~\cite{Liu2010} findings for Eu and Yb. Notably, our results show a slight increase in magnitude from Tm to Ho and, finally, to Dy, suggesting variations in the interaction strength with the graphene substrate.

In the following, 
we focus on the H-site and perform structural relaxations allowing the $4f$ atom 
to adjust its height along the c-axis ($z$-direction with respect to the graphene) until reaching minimization of total energy and forces acting on the RE atom,
with the SOC included self-consistently. The obtained ground state properties including the $f$ and $d$ occupations, spin and orbital moments are provided in Table~\ref{table2}.
\begin{table*}[!t]
%\centering
\caption{\label{table2} Perpendicular distance from the graphene monolayer (\AA), $d$ and $f$ occupation, magnetic moment and orbital moment ($\mu_\text{B}$) of the rare-earths atoms on graphene in the H-site. Calculations have been performed in presence of SOC.}
%\begin{indented}
\begin{tabular*}{\linewidth}{c@{\extracolsep{\fill}}cccccc}
\midrule[0.4mm]
Rare-earth & distance $d_0$ (\AA) & $d_\mathrm{occ}$ &  $f_\mathrm{occ}$& $m_s^\mathrm{RE}$ ($\mu_\text{B}$) & $m_l^\mathrm{RE}$ ($\mu_\text{B}$) & $m_s^\mathrm{tot}$ ($\mu_\text{B}$)\\ 
\midrule[0.2mm]
\bfseries{Dy} &$2.49$& $0.262$ & $9.891$ & $4.040$ & $5.876$ &$4.174$\\ 
\bfseries{Ho} & $2.50$& $0.250$ & $10.881$ & $3.045$ & $5.905$& $3.150$\\ 
\bfseries{Tm} &$2.47$& $0.237$ & $12.867$ & $1.027$ & $3.000$& $1.072$\\ 
\midrule[0.4mm]
\end{tabular*}
%\end{indented}
\end{table*}
\begin{figure*}[t!]
\centering
    \includegraphics[width=\textwidth]{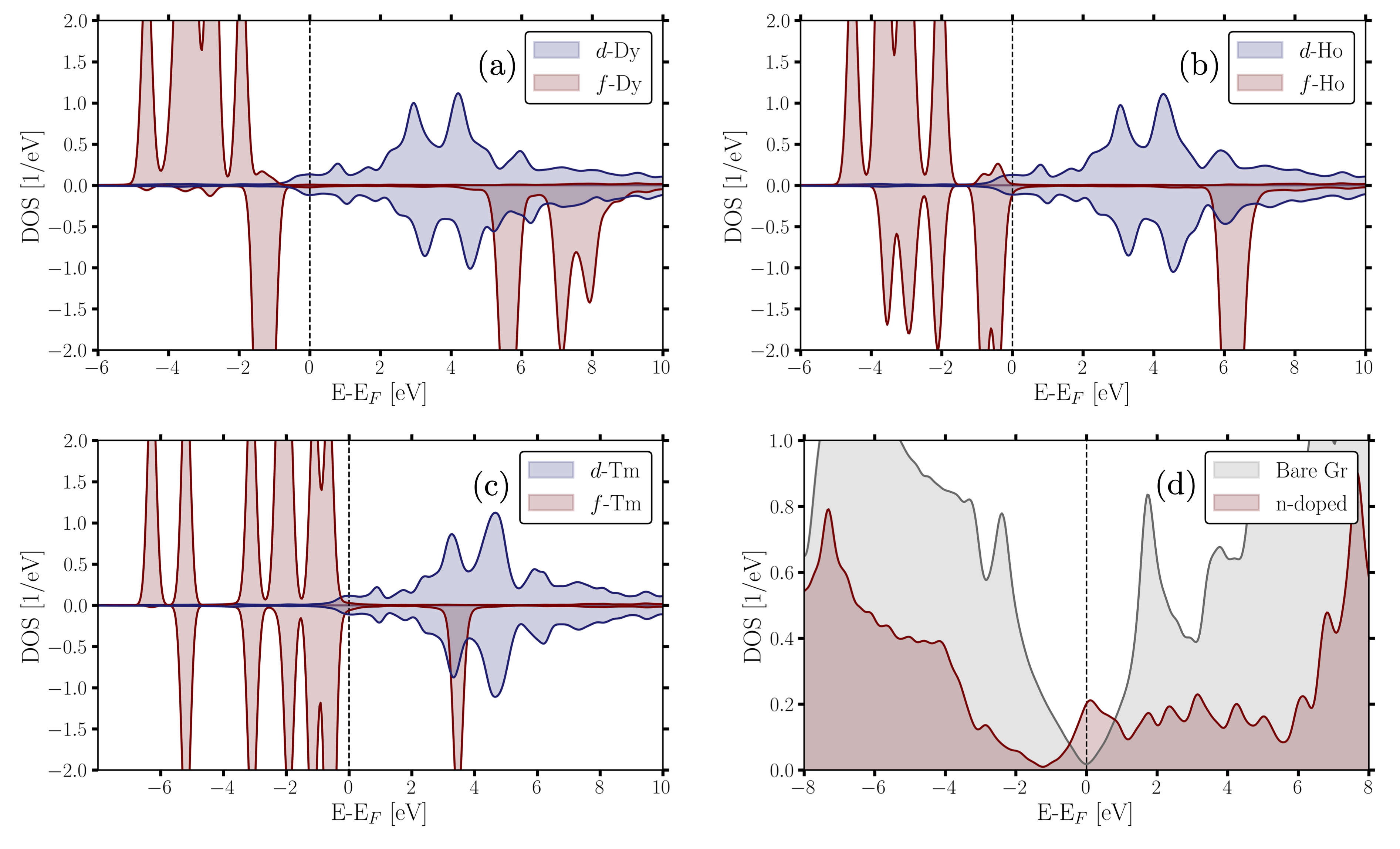}
    \caption{Spin-polarized density of states of the $d$ (blue) and $f$ (red) electrons of (a) Dy (b) Ho and (c) Tm, on top of graphene. The upper half of the plots displays the majority states, while the lower panel is relative to the minority states. The value $E-E_F=0$ corresponds to the Fermi energy. (d) DOS of $n$-doped graphene (shown is the contribution of the carbon atoms) in the Ho/Gr system (red) and DOS of bare graphene (grey). All calculations where 
    performed including SOC self-consistently.}
    \label{DOS}
\end{figure*}

\subsection{Electronic Properties}
The $5d$ and $4f$ occupations shown in Table~\ref{table2} clearly show that rare-earth atoms 
undergo a semi-atomic like behavior, where the $4f$-shell follows Hund's rules while acquiring 
some $d$ occupation. This semi-atomic picture can be visualized in the spin-resolved density of 
states (DOS) shown in Figs.~\ref{DOS}(a-c), where the $f$ occupation is shown in red and the $d$ occupation
in blue. The DOS of the pristine graphene is given in Fig.~\ref{DOS}(d), and it is compared to the doped 
graphene DOS. We only show Ho/Gr as it looks identical for the Dy and Tm impurities.
Fig.~\ref{DOS}(d) shows that the RE/Gr systems exhibit a metallic behavior due to the $n$-doping 
coming from the lanthanide impurities. The magnitude of the doping can be estimated from the energy
difference between the pristine graphene and the doped one, which is of the order of $1.4$ eV,
in agreement with the charge transfer calculations carried out in Ref~\cite{Basiuk2022}. This charge 
transfer is driven by the hybridization between the graphene $p_{z}$ orbitals and the magnetic 
atom's $d$ orbitals. 

A clear correlation between the adsorption energies in the H-site and the $d$ 
occupation appears with Dy having the highest $d$ occupation (Table~\ref{table2}) and showing also 
the strongest bonding towards the substrate (Table~\ref{table1}). Moreover, we observe that the small  
bonding energy at the T and B sites discussed previously is reflected in a low $d$ occupation of 
the rare-earth atoms, indicating the major role played by the $d$-electrons in the chemical bonding.
The LDOS of the $4f$ atoms, depicted in Fig.~\ref{DOS}(a-c), shows that the $f$-states are hybridized
and spread over a large energy window. These states exhibit an insulating character featuring a gap between the
occupied and unoccupied states, with the occupied states lying close to the Fermi energy. 

\subsection{Magnetic Moments}
 The orbital ($m_l^\mathrm{RE}$) and spin ($m_s^\mathrm{RE}$) moments of the REs provided in Table~\ref{table2} follow closely Hund's rules. Nonetheless, 
the $m_s^\mathrm{RE}$ values are slightly higher than the \rm{RE}$^{2+}$ ionic atom case due to the spin polarization of
$d$-occupation via an intra-atomic $f-d$ exchange interaction. This $d$ spin polarization is 
about $0.04$ $\mu_\text{B}$ for Dy, $0.03$ $\mu_\text{B}$ for Ho and $0.01$ $\mu_\text{B}$ for Tm, respectively. This decay reflects the decreasing value of $m_s^\mathrm{RE}$ from Dy to Tm. The muffin-tin C spin polarization has a very small induced moment of $\sim 0.001$ $\mu_\text{B}$. The total magnetic moment of the system $m_s^\mathrm{tot}$ is naturally dictated by the $4f$, which is inversely proportional to the $f$ occupation for late-series REs as shown in Table~\ref{table2}. The interstitial region of the structure (space between the atoms) is mostly represented by the delocalized $\pi$-orbitals of graphene and $d$ electrons of the RE atom. These $\pi$ and $d$ electrons carry a small spin moment $\Delta m=m_s^\mathrm{tot}-m_s^\mathrm{RE}$ induced by the presence of the RE atom. The induced magnetization is proportional to the RE's atom spin moment and is given by $0.125$ $\mu_\text{B}$ for Dy/Gr, $0.094$ $\mu_\text{B}$ for Ho/Gr, and $0.036$ $\mu_\text{B}$ for Tm/Gr.

\section{Magnetic anisotropy and crystal field coefficients}
\label{MAE_curves}
\begin{figure*}[t]
\centering
    \includegraphics[width=\textwidth]{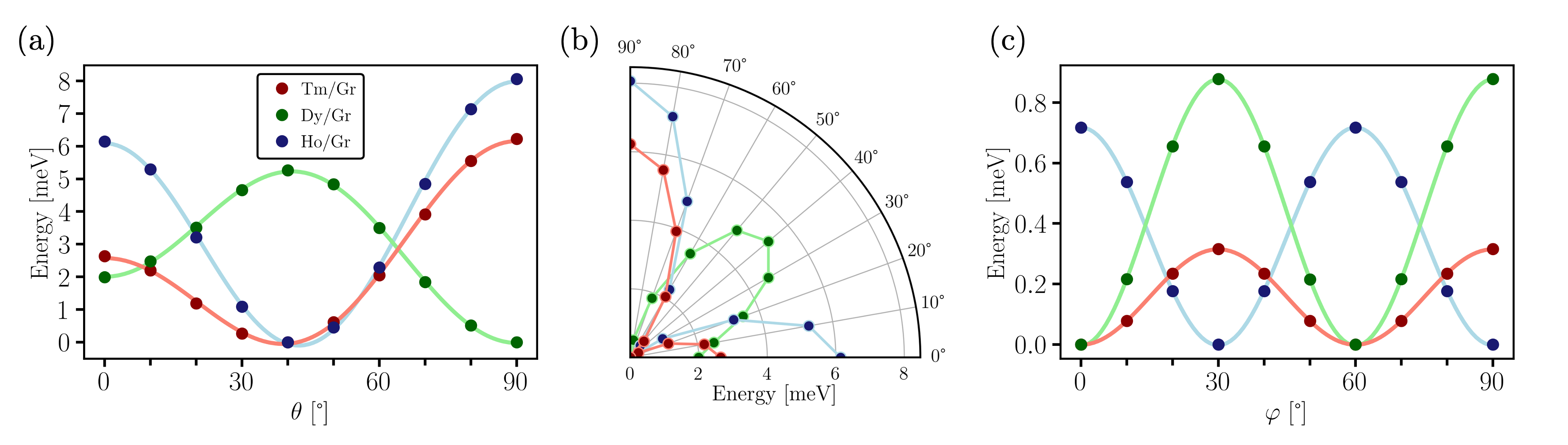}
    \caption{(a) Out-of-plane $\theta=0^\circ$ to in-plane $\theta=90^\circ$  magnetic anisotropy energy curves for Dy (green), Ho (blue) 
    and Tm (red) on graphene: the total energy is plotted as function of the polar angle. (b) Alternative 
    representation of the DFT+$U$ data in a polar plot for the computed systems. (c)  In-plane 
    magnetic anisotropy energy curves for Dy, Ho and Tm on graphene: the total energy is plotted as function of the azimuthal angle $\varphi$. Full dots indicate the DFT+$U$ data, while the full lines display the fitting 
    curves.}
    \label{MAE_curves}
\end{figure*}
The adsorption of an atom on a surface leads to physical properties that
drastically differ from the isolated atom. As the RE atoms are in contact 
with the surface they experience the electric field produced by the surrounding 
atomic charges. This crystal 
field~\cite{hutchings1964point} (CF) results in a lowering of the symmetry 
of the spherical potential in an isolated atom.
The form of this field is dictated by the lattice symmetry and local chemical environment and determines the angular dependence of the total energy upon rotation of the magnetization, \textit{i.e.}, the magnetic anisotropy. The conventional anisotropy energy functional describing the angular dependence of the magnetic anisotropy energy for a hexagonal system reads as~\cite{skomski2020anisotropy}: 
\begin{equation}
\begin{split}
E_{an}=&K_{1}\sin^{2}\theta+K_{2}\sin^{4}\theta+K_{3}\sin^{6}\theta+\\
&+K_{4}\sin^{6}\theta\cos6\varphi\quad.
\label{MAE}
\end{split}
\end{equation}
$K_{i}$ are the magnetic anisotropy constants, $\theta$ is the polar angle 
between the magnetization and $z$-axis, while $\varphi$ is the azimuthal angle 
between the magnetization and the $x$-axis.
In the following, we compute the anisotropy constants $K_{i}$ by fitting 
the changes in the total energy upon rotation of the magnetic moment relative to the crystal lattice. The results are given in Fig.~\ref{MAE_curves} and are obtained self-consistently with 
DFT+$U$ and SOC as discussed in Sec.~\ref{parameters}.
The dotted data represent the \textit{ab initio} results which are then fitted 
with the continuous lines using Eq.~\eqref{MAE}.

Fig.~\ref{MAE_curves}(a) displays the total energy change 
within the ($xz$)-plane by steps of $10^\circ$. An alternative visualization of the out-of-plane curves in terms of a polar plot is given in panel (b). 
Fig.~\ref{MAE_curves}(c) displays the basal anisotropy within the graphene plane 
with an azimuthal rotation angle $\varphi$ away from the $x$-axis. The fitted values 
of $K_i$ are summarized in Table~\ref{table3} for the investigated systems. 
The lowest order {constants} are an order of magnitude bigger than the third one. 
The in-plane {constants} $K_4$ are again one order of magnitude weaker than $K_3$ with 
exception of Tm. In all three 
systems, the magnetic anisotropy energy is dominated by the $K_1$ and $K_{2}$ constants 
and deviates from $\sin^{2}\theta$, indicating the crucial role of higher order 
anisotropies in an open $4f$-shell.

The green curve in Fig.~\ref{MAE_curves} shows the anisotropy energy for Dy/Gr, for which an in-plane easy-axis is obtained ($\theta=90^{\circ},\ \varphi=0^{\circ}$). The energy 
difference between the easy and the $z$-axis is $\Delta E \simeq 2$ 
meV, while the energy barrier to overcome to switch the magnetization is about $5.3$ meV. Ho/Gr (blue curve) has an intermediate easy-axis with 
the configuration ($\theta=42.67^{\circ},\ \varphi=30^{\circ}$) indicating a 
canted magnetization with respect to the graphene sheet. The positive sign of $K_4$ leads to a non-zero basal angle $\varphi=30^{\circ}$ for the minimal energy.

In Tm/Gr (red curve) the magnetic anisotropy curve is qualitatively 
similar to Ho/Gr with lower energy barriers and the ground state corresponds to a tilted
magnetic configuration ($\theta=39.08^{\circ}$,\ $\varphi=0^{\circ}$). 
The direction of the easy-axis for each system can be explained by examining
the values $K_i$. The overall shape of the energy curves can be derived by calculating the second derivative of Eq.~\eqref{MAE} neglecting $K_3$ and $K_4$ and considering $\sin^2\theta=-K_1/2K_2$, which leads to $\frac{\partial^2 E_{an}}{\partial \theta^2}=-2K_1\left(\frac{2K_2+K_1}{K_2}\right)$. For all the RE/Gr systems, the term in parenthesis is positive and thus the behavior is fully determined by $K_1$, giving rise to an energy valley if $K_1<0$ or an energy hill if $K_1>0$. Computing $\frac{\partial^2 E_{an}}{\partial \theta^2}$ for $\theta=0^{\circ},90^{\circ}$ permits to determine the behavior at the extrema.
For systems with $K_1>0$, such as Dy/Gr, the curve at $\theta=0$ exhibits a convex trend, whereas for Ho/Gr and Tm/Gr, with $K_1<0$, the curve shows a concave shape. Similarly, the curvature at $\theta=90^{\circ}$ is governed by $\frac{\partial^2 E_{an}}{\partial \theta^2}|_{\theta=90^{\circ}}=-2K_1-4K_2$, producing a concave trend for Ho/Gr and Tm/Gr  and a convex shape for Dy/Gr. The in-plane curves in Fig.~\ref{MAE_curves}(c), reflect the in-plane six-fold ($C_{6v}$) symmetry. The functional form of the energy is therefore 
$\propto\cos6\varphi$. The amplitude of the oscillation is highest for Dy, followed by Ho and Tm in accordance with the values of $K_4$ given in Table~\ref{table3}.
\begin{table}
\centering
\caption{\label{table3} Magnetic anisotropy constants obtained via fitting of DFT+$U$ data depicted in Fig.~\ref{MAE_curves} for Dy, Ho and Tm on graphene. Units are in meV.}

\begin{tabular*}{\linewidth}{@{\extracolsep{\fill}}
        l
    *{5}{S[detect-weight,   % <--
           mode=text,       % <--
           table-format=1.3]}}
\midrule[0.4mm]
 \mc{RE/Gr} &\mc{$K_1$} & \mc{$K_2$} & \mc{$K_3$} & \mc{$K_4$}\\
\midrule[0.2mm]
\B{Dy/Gr}&15.355&-18.918&1.536&-0.441  \\ 
\B{Ho/Gr}& -27.734&32.218&-2.591&0.360   \\
\B{Tm/Gr} &-13.285&16.720&0.146&-0.158   \\
\midrule[0.4mm]
\end{tabular*}
\end{table}
Nevertheless, the classical formulation described above does not take into account effects at the quantum level that favour magnetization reversal, thus we proceed with a quantum-mechanical description.
The CF indeed splits the $(2J+1)$-fold degeneracy of an isolated atom into 
sub-levels~\cite{jensen1991rare}. For the adatoms, the actual form of the CF 
Hamiltonian depends on the specific point-group of the adsorption site.
Focusing 
on the H-site of graphene which has a $C_{6v}$ symmetry, the crystal field 
Hamiltonian reads~\cite{kuzmin2007chapter,radwanski1989magnetocrystalline}: 
\begin{equation}
    \hat{H}_{\text{CF}}=C_{2}^{0}\hat{O}_{2}^{0}+C_{4}^{0}\hat{O}_{4}^{0}+C_{6}^{0}
    \hat{O}_{6}^{0}+C_{6}^{6}\hat{O}_{6}^{6}\quad,
    \label{CF}
\end{equation}
where $C_{l}^{m}$ are the CFP, and the $\hat{O}_{l}^{m}$ are the Stevens operators~\cite{hutchings1964point,jensen1991rare}, given in Appendix \ref{appendix:a}. $l$ and $m$ represent the angular and magnetic quantum numbers respectively, which arise from the formulation of the crystal field potential in terms of spherical harmonics. These spherical harmonics are then converted in Stevens operators adopting the Stevens' operator-equivalent method~\cite{stevens1952matrix}. 
The operators of Eq.~\eqref{CF} 
act on the atomic $J_z$ eigenstates removing their degeneracy and mixing the different magnetic states.
The first $3$-terms contain powers of $J$ and $J_z$ which split the $J_{z}$ states generating a specific energy  landscape of the quantum levels depending on the CFP magnitude and determine the difference between the 
highest and lowest states, \textit{i.e.}\ the energy barrier to overcome in order to 
reverse the magnetization. The last term in Eq.~\eqref{CF} contains the ladder operators 
$\hat{J}_{\pm}=\hat{J}_x\pm i\hat{J}_y$ in the form $\hat{O}_6^6=1/2(\hat{J}_+^6+\hat{J}_-^6)$, which 
mixes the $J_{z}$ states differing by $\Delta J_{z}=\pm6,\pm12$, and can
thus possibly generate tunnel-split doublets (symmetric and antisymmetric 
linear combinations) with quenched $\braket{J_{z}}$ value that can 
significantly reduce the energy barrier for a spin-flip event inducing quantum tunneling of magnetization.
\begin{table*}[t!]
\centering
\caption{\label{table4}Crystal field coefficients obtained via equation~\ref{eqK} for Dy, Ho and Tm on graphene. Results are shown in meV.}

\begin{tabular}{@{}SSSSS} 
\midrule[0.4mm]
 &\mc{$C_{2}^{0}$}&\mc{$C_{4}^{0}$}&\mc{$C_{6}^{0}$}&\mc{$C_{6}^{6}$}\\
\midrule[0.2mm]
\bfseries{Dy/Gr}& 0.025 & -1.717$\cdot$ $10^{-4}$ & -7.381$\cdot$ $10^{-8}$ & -4.895$\cdot$ $10^{-6}$ \\

\bfseries{Ho/Gr} & -0.039 & 3.904$\cdot$ $10^{-4}$ & 1.992$\cdot$ $10^{-7}$ & 6.394$\cdot$ $10^{-6}$    \\

\bfseries{Tm/Gr} & -0.190 &  9.229$\cdot$ $10^{-3}$  & -8.026$\cdot$ $10^{-6}$ & -2.006$\cdot$ $10^{-3}$ \\
\midrule[0.4mm]
\end{tabular}
\end{table*}
Indeed, when quantum states at 
$\braket{J_{z}}=0$ are formed, the system does not necessarily need 
to overcome the whole energy barrier extending from the ground-state
to the highest lying state in order to exhibit magnetization reversal,
but can tunnel through this barrier towards the opposite magnetization
state for example via thermal excitation.
In this picture, the CFP determine how the $\hat{O}_l^m$ split the $J_z$ states 
and are thus an essential ingredient to understand the mechanisms which determine 
the magnetic stability of single-atom magnets.
Therefore, the knowledge of 
the CFP is highly demanded and helps identifying possible systems protected 
from magnetization-reversal events that could be an appealing choice for
stable magnetic units. 
In the following, we proceed by providing a simple approach 
to evaluate the CFP: using 
first order perturbation theory we compute the classical CF energy~\cite{Patrick2020}. 
Assuming that the MAE contributions come fully from the $4f$-orbitals,
we then extract the CF coefficients as linear combinations of the $K_{i}$ constants obtained from the above discussed fitting.
In the limit where the CF effects are small in comparison to the magnetic exchange 
field, one can focus on the CF contribution to the atomic Hamiltonian and apply 
first order perturbation theory. The energy change $E_{\text{CF}}$ attributed to 
$\hat{H}_{\text{CF}}$ reads then~\cite{Yamada}:
\begin{equation}
\begin{split}
E_{\text{CF}} &= \sum_{l=2,4,6}C_{l}^{0}
\ev{\hat{O}_{l}^{0}}+C_{6}^{6}\ev{\hat{O}_{6}^{6}}\quad,\\
\ev{\hat{O}_{l}^{m}} & =\braket{J,M|\hat{O}_l^m|J,M}\quad\\
&= f_l(J)F_l^m(\theta)G_m(\varphi)\quad.
\end{split}    
\label{CF_energ}
\end{equation}
$M=+J$ for heavy REs and $f_l=2^{-l}(2J)!/(2J-l)!$. The angular functions 
$F_l^m(\theta)$ and $G_m(\theta)$ are determined in Ref.~\cite{Yamada} and 
listed in appendix~\ref{Appendix:b}. By equating Eq.~\eqref{MAE} and 
Eq.~\eqref{CF_energ}, we obtain a linear relation between the CFP and 
the $K_{i}$ coefficients of the MAE: 
\begin{equation}
    \begin{split}
    K_{1}&=-3f_{2}C_{2}^{0}-40f_{4}C_{4}^{0}-168f_{6}C_{6}^{0}\quad,\\
    K_{2}&=35f_{4}C_{4}^{0}+378f_{6}C_{6}^{0}\quad,\\
    K_{3}&=-231f_{6}C_{6}^{0}\quad,\\
    K_{4}&=f_{6}C_{6}^{6}\quad.
    \end{split}
    \label{eqK}
\end{equation}
Using the coefficients $K_{i}$ and Eq.~\eqref{eqK}, the CFP can be determined 
for the ground state configurations for Dy ($J=8$), Ho ($J=15/2$) and for Tm ($J=7/2$). The results are summarized in Table~\ref{table4}. The sign of 
$C_{2}^{0}$ determines the orientation the parabolic dispersion around
$J_{z}=0$ since it multiplies $\hat{O}_{2}^{0}=3\hat{J}_z^2-J(J+1)$, thus it determines the easy-axis 
of the system considering a first order anisotropy, \textit{i.e.}, when only $K_1$ is non-zero. 
A negative $C_{2}^{0}$ 
corresponds then to an out-of-plane easy-axis ($K_1>0$), while a positive $C_{2}^{0}$ is associated 
to an in-plane easy-axis ($K_1<0$). 
\begin{figure}[b!]
\centering
    \hspace{-5mm}
    \includegraphics[scale=0.13]{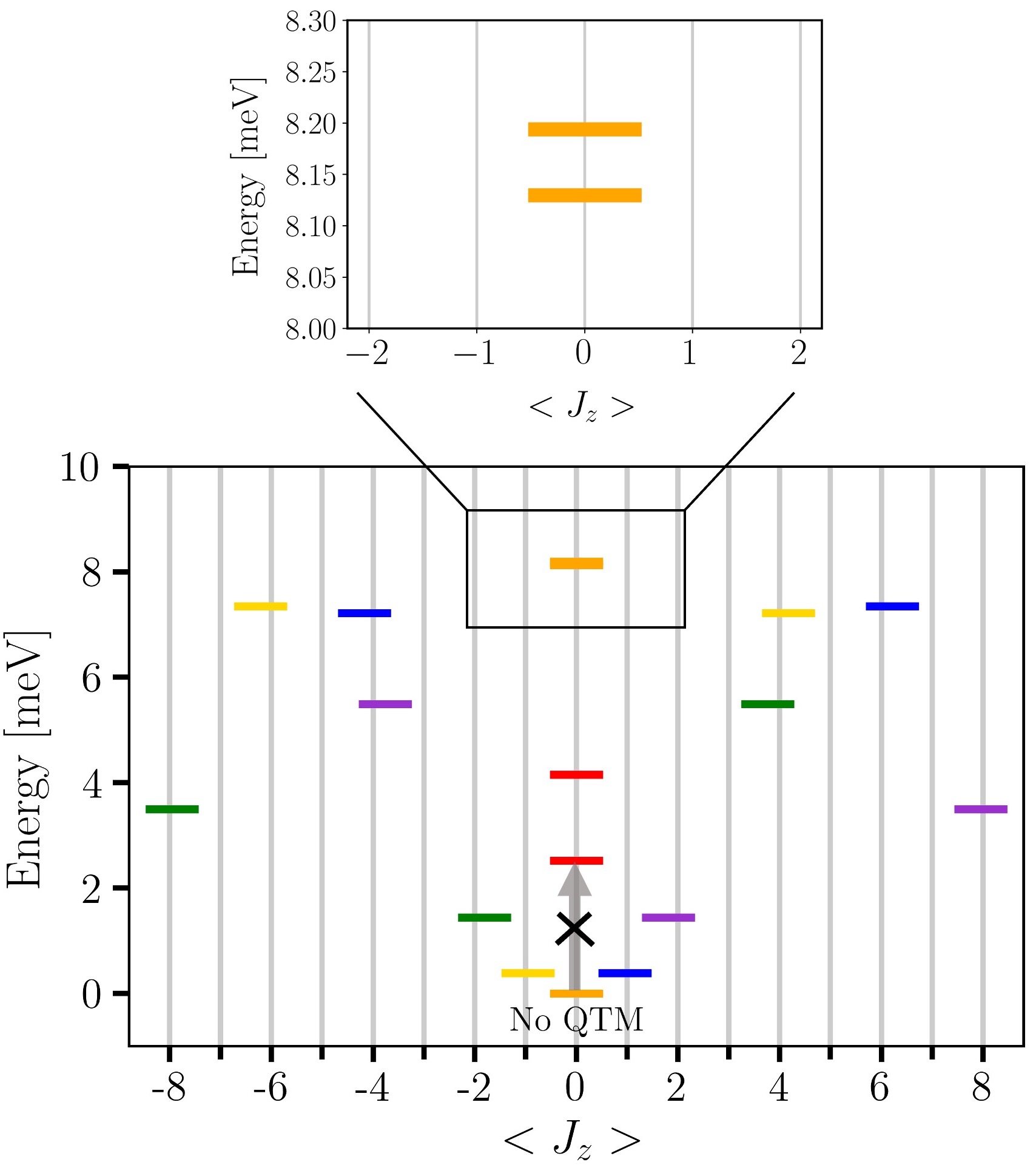}
    \caption{Multiplet splitting of Dy/Gr adopting the CFP values obtained from reverse-engineering via the magnetic anisotropy constants. States in the same color correspond to linear combinations of $\ket{J_z}$ differing by $\Delta J_z=\pm6,\pm12$. Inset shows the $\sim50\mu$eV energy splitting of the $\ket{J_z=-6}$, $\ket{J_z=0}$, $\ket{J_z=+6}$ doublet.}
    \label{Dy_splitting}
\end{figure}

Considering Dy/Gr with integer $J=8$, we find the $C_2^{0}$ 
coefficient is positive leading to a single ground state with $\braket{J_z}=0$,
as evidenced in the multiplet splitting shown in Fig.~\ref{Dy_splitting}. Here, states with the same color represent a mixture of different states differing by $\Delta J_z =\pm 6,\pm 12$, and therefore can slightly deviate from the expectation values $\braket{J_z}$ of pure states. The strength of this mixing is dictated by the value of the $C_6^6$ parameter. In particular, for Dy/Gr, several of these linear combinations form and of these two sets - one being a mixture of $\ket{J_z=-3}$ and $\ket{J_z=+3}$ (shown in red) and one of $\ket{J_z=-6}$, $\ket{J_z=0}$, $\ket{J_z=+6}$ (orange, the state at $\sim8$ meV corresponds to a doublet, better resolved in the inset of Fig.~\ref{Dy_splitting}) - appear at quenched $\braket{J_z}=0$ value. These kind of states in general can compromise the stability of the magnetization inducing quantum tunneling of magnetization. Nevertheless, in the present case there is a single energy minimum and magnetization reversal is not possible.

We compare the values for Dy/Gr with Ref.~\cite{SHICK2019} where the CFP are 
reported in the standard notation $A_{l}^{m}\braket{r^l}=C_{l}^{m}/\theta_l(J)$, with 
$\theta_l(J)$ the Steven's factors for a total angular momentum $J$. 
For the Dy with the configuration $J=8$ (for Dy$^{2+}$), the Steven's 
factors are $(\theta_2,\theta_4,\theta_6)=(-0.222\cdot10^{-2},-0.333
\cdot10^{-4},-1.3\cdot10^{-6})$ (these are the values associated Ho$^{3+}$). 
The resulting values are $(A_2^{0},A_4^{0},A_6^{0},A_6^{6})=(-11.287,5.156,0.057,3.765)$ meV. The largest coefficients namely the $A_2^{0}$ and $A_4^{0}$ are in good agreement
with the values obtained in Ref.~\cite{SHICK2019} using the Hubbard-I 
approximation, while deviations in magnitude are observed in $A_6^{0}$ 
and $A_6^{6}$. These different values might be a consequence of the
supporting Ir substrate included in~\cite{SHICK2019}.

The energy required to overcome the energy barrier from the lowest to the highest lying $J_z$ state in the quantum picture is associated to the classical first order magnetic anisotropy, \textit{i.e.}, the energy involved in the magnetization reversal from out-of-plane to in-plane. Comparing the quantum and classical models, it can be seen that the quantum approach corresponds qualitatively to the classical magnetization rotation, with a in-plane magnetic ground-state for Dy/Gr.

Concerning Ho/Gr and Tm/Gr systems, the CFP lead to a degenerate ground-state with non-minimal $\braket{J_z}$  (Fig.~\ref{Ho/Tm_splitting}(a) and (b) in Appendix~\ref{Appendix:e} respectively). The energy trend favours a canted magnetic ground-state as determined in the MAE curves. Both magnetic atoms are characterized by a half integer $J$ value and are protected against the formation of states at $\braket{J_z}=0$ by Kramer's degeneracy and consequently against quantum tunneling of magnetization via those states, such that in principle the system has to overcome the whole energy barrier from the lowest multiplet to the highest lying multiplet in order to exhibit spin-flip. Nonetheless, based on the values of the higher order crystal field coefficients, the dispersion of the 
$J_z$ states can have different non-monotonic shapes in which a faster 
way for a spin-flip event might be favoured, for example first order 
transitions at finite temperature via phonon or electron scattering 
events, that can be followed by quantum tunneling of magnetization.

\section{Temperature effects}
In this section we discuss the effect of temperature on magnetisation reversal or stability. We distinguish two mechanisms in two different temperature regimes. At finite temperatures, magnetization reversal can occur through thermal activation, enabling the system to overcome the minimal energy barrier when the temperature is sufficiently high. This leads to an Arrhenius-like relationship for the magnetic lifetime under the condition that no external magnetic field is present. 
At lower temperatures, although the thermal energy may not be sufficient to overcome the energy barrier, it can prompt excitations to metastable higher-energy states that enable thermally assisted quantum tunneling of the magnetization \cite{gatteschi2003quantum}. This phenomenon involves scattering processes, such as interactions with substrate phonons. The mathematical representation employs operators $\hat{J}_z$, $\hat{J}_+$, and $\hat{J}_-$, enabling transitions between states characterized by angular momentum changes of $\Delta J_z=0,\pm 1$. Hence, in this first-order perturbation scenario, the operator $\hat{O}_6^6$ facilitates the coupling between states of equal energy having angular momentum differences of $\Delta J_z= 0 \pm 6k, -1 \pm 6k, 1\pm 6k$. Here, the parameter $k$ assumes integer values depending on $J$~\cite{Baltic2018}.
This process thus involves the transition to a higher-energy state, from which subsequent quantum tunneling can take place.

Within the scope of the systems under investigation, the Dy/Gr system does not display a magnetic bistability. Instead, it maintains a singular non-degenerate ground state at $\braket{J_z}=0$,  consequently ruling out the possibility of magnetization reversal.
However, when considering the half-integer spin system Ho/Gr, the two degenerate ground states manifest at $\braket{J_z}=\pm 11/2$. The energy barrier separating these two ground states is substantial, approximately $\sim 14$ meV, which corresponds to an activation barrier of $U=162$~K in the relaxation time $\tau\propto e^{U/\text{k}_\text{B}T}$, where $\text{k}_\text{B}$ is the Boltzmann constant and $T$ the temperature.
\begin{figure*}[t!]
\centering
    \includegraphics[scale=0.175]{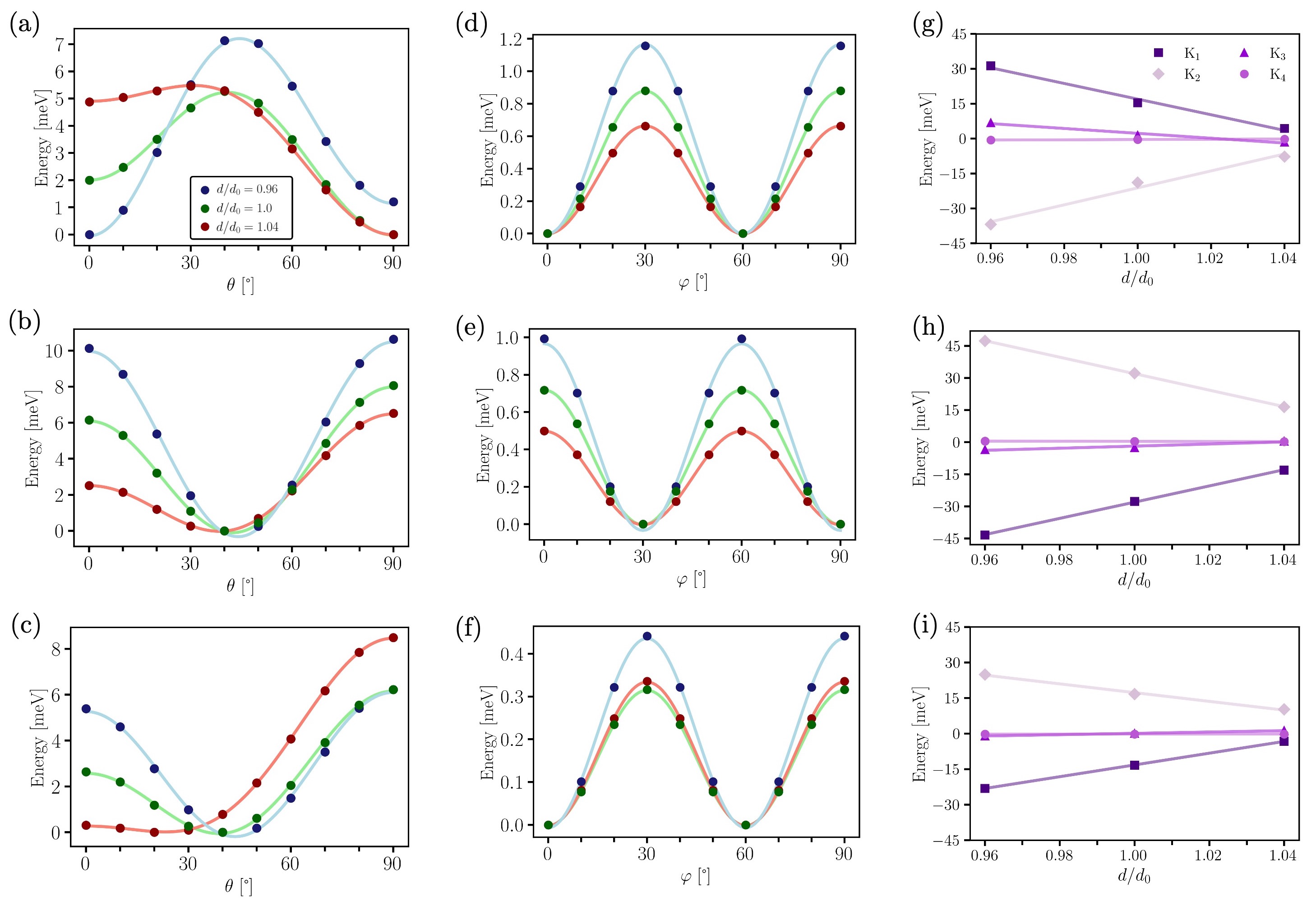}
    \caption{MAE curves (out-of-plane and in-plane) for different distances, namely 
    $d/d_{0}=0.96, 1.0, 1.04$ (blue, green and red respectively) of the rare earth 
    adatoms from the graphene monolayer. (a) and (d) correspond to the out-of-plane 
    and in-plane curves of Dy/Gr; (b) and (e) correspond to the out-of-plane and 
    in-plane curves of Ho/Gr; (c) and (f) correspond to the out-of-plane and in-plane 
    curves of Tm/Gr. Note the different scales for the energies of the out-of-plane and in-plane results. For each system, the last column (Figures (g)-(i)) shows the 
    respective magnetic anisotropy constants $K_i$ ($i=1,2,3,4)$ obtained via the 
    fitting of the MAE curves. Specifically, (g) shows the $K_i$ for Dy/Gr, (h) for 
    Ho/Gr and (i) for Tm/Gr. Points correspond to DFT+$U$ data while lines to the 
    fitting curves.}
    \label{Magnetoelastic}
\end{figure*}
However, interactions with substrate phonons can establish a connection between these two states and the closest accessible states via thermal excitation. Specifically, these accessible states are characterized by $\braket{J_z}=\pm 13/2$ and are positioned at an energy gap of roughly $\Delta E \sim 1.8$~meV ($21$~K), from which assisted quantum tunneling is possible.
Shifting focus to the Tm/Gr system, the doubly degenerate ground states possess an expectation value of $\braket{J_z}\sim\pm 5/2$. Overcoming the entire energy barrier separating these ground states would require an energy of $209$~K (equivalent to $18$~meV). Furthermore, there exists an energy gap of $95$~K (approximately $\sim 8.2$~meV) to the first excited state at $\braket{J_z}=\pm 3/2$, which is inherently protected against quantum tunneling.

\section{Magnetoelastic coupling}
\label{magnetoel_section}
The particularly large MAE found in these materials is a consequence of the 
localized and partially filled $4f$-shells together with the surrounding crystal field 
of the graphene substrate. In the case of a half-filled $4f$-shell with a vanishing total 
orbital moment $\bf{L}$ (Eu and Gd), the SOC contribution of the $4f$-electrons is tiny and
hence the MAE is drastically reduced \cite{Eu_michely,Gd_orb}.  The different values of $\bf{L}$ correspond to specific shapes of the charge cloud \cite{jiang2015prediction,sievers1982asphericity} (see Fig.~\ref{Dy_spindensity}) that interact 
with the neighboring sites as the spin moment $\bf{S}$ rotates. Given the strong dependence of the 
MAE on the shape of $4f$-charge distribution, strong changes in the MAE can occur 
due to mechanical deformations. The induced strain might for 
instance induce a displacement of the charge density inside the structure and through 
SOC effect lead to change the orientation of $\bf{S}$.
Here, we simulate the effect of strain of MAE by changing the height of the rare-earth 
atoms with respect to the graphene sheet and analyze magneto-elastic coupling. From an experimental perspective, this shift in distance can be realized, for instance, by modifying the chemical reactivity or the charge state of the graphene sheet through intercalation of dopands between graphene and the substrate~\cite{kraus2022selecting,huttmann2015tuning,schumacher2013backside}.
Figs.~\ref{Magnetoelastic}(a-c) show the change of the out-of-plane MAE, while 
Figs~\ref{Magnetoelastic}.(d-f) depict the in-plane MAE. Three different distances 
were considered, $d/d_{0}=\{0.96, 1.0, 1.04\}$, where $d/d_{0}=1.0$ represents the 
initial relaxed position of the adatom, $d$ being the shifted height and $d_0$ the equilibrium height. 
The MAE is once more 
obtained from total energy self-consistent calculations (dotted data) and fitted with Eq.~\eqref{MAE} (continuous line). 

First, we focus on the out-of-plane MAE which increases, in terms of modulus of the $K_i$ constants, as the adatom is compressed towards 
the surface for all cases. 
The dependence of the MAE constants $K_{i}$ as a function of 
the distance is shown in Figs.~\ref{Magnetoelastic}(g-i), where the systematic increase of the 
$K_{i}$ is due to the enhancement of the crystal field as the impurity gets closer to 
the substrate. We also note that the complex shapes once more require more coefficients,
$K_2$ and $K_3$, which result in canted minimum energy solutions for Ho and Tm.
As discussed also in Section~\ref{MAE_curves}, the sign of $K_1$ is reflected in 
the generation of an energy hill or valley in the MAE curve, that are more pronounced 
for higher absolute values of $K_1$. 
In general, $K_2$ exhibits an opposite sign when comparing Dy/Gr to Ho and Tm/Gr
and a slightly bigger absolute value than $K_1$ for all the systems. 
The contribution of $K_2$ leads to a tilted easy(hard)-axis for Ho, Tm (Dy).
Also, $K_3$ exhibits 
opposite sign when comparing Dy/Gr with Ho/Gr and Tm/Gr. Nevertheless, the major difference shows up in 
the module of this constant, which assumes a higher value in the case of Dy ($6.821$ \ meV 
compared to $-3.434$ \ meV and $-1.032$ \ meV for Ho/Gr and Tm/Gr respectively) at $d/d_0=0.96$, 
and hence has larger influence on the MAE curve, causing a change in the easy-axis from in-plane for $d/d_0=1.0$ and $d/d_0=1.04$ to out-of-plane for $d=0.96$.

The in-plane anisotropies in Figs.~\ref{Magnetoelastic}(d-f) display similar behavior as discussed
in Section~\ref{MAE_curves} with a periodicity of $\varphi=60^{\circ}$. The amplitudes of 
the oscillations are given in terms of $K_4$, and, similarly to the out-of-plane 
coefficients, it is enhanced by the reducing $d/d_{0}$. The only exception appears 
for Tm/Gr, for which $d/d_{0}=1.04$ seems slightly larger that $d/d_{0}=1.0$. A more detailed analysis for the latter is reported in Appendix~\ref{Appendix:c}. 

\begin{figure}[h!]
\centering
\hspace{-6mm}
    \includegraphics[scale=0.15]{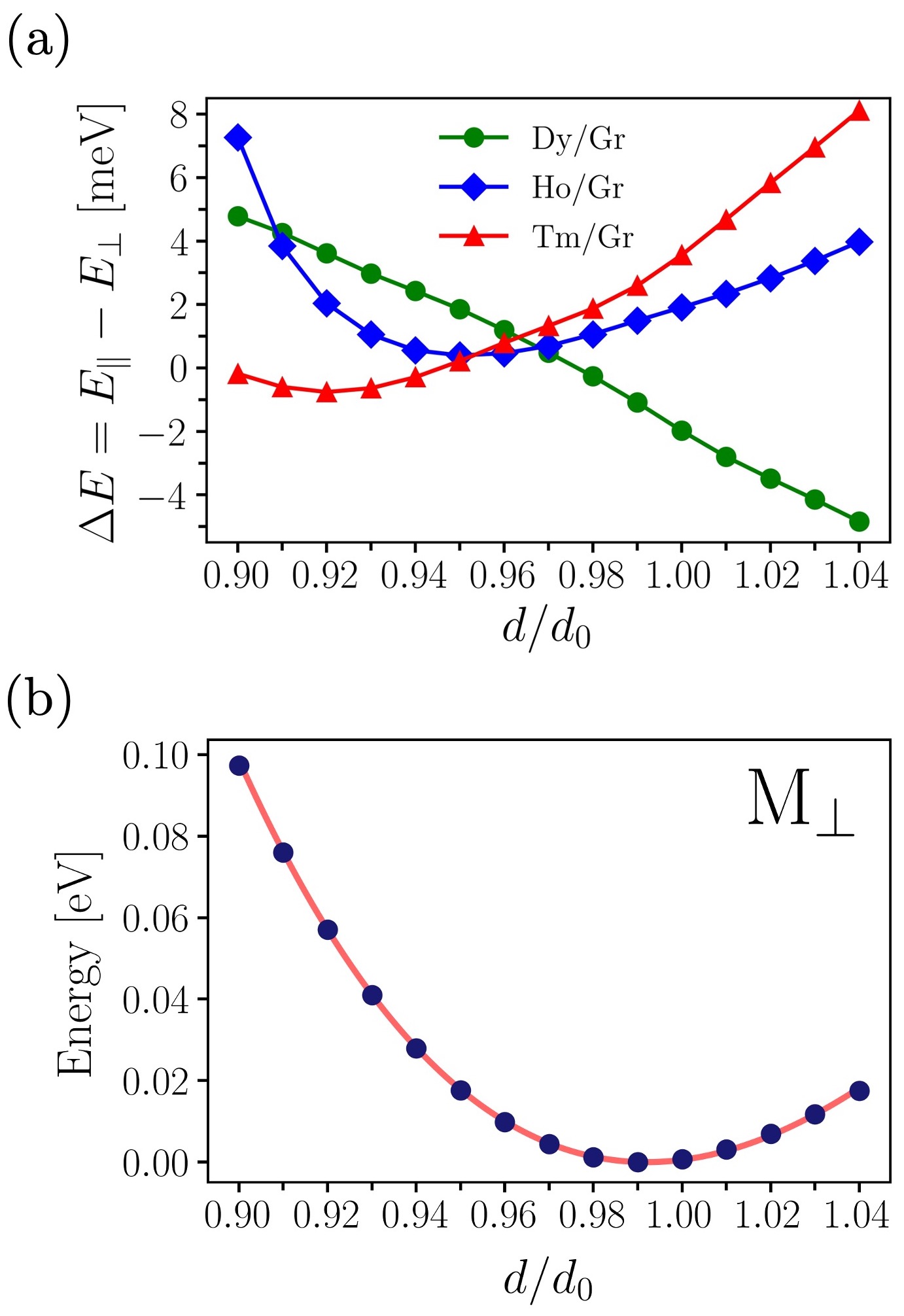}
    \caption{(a) Magnetic anisotropy energy as a function of the distance of the RE with respect to graphene for Dy (green), Ho (blue) and Tm (red) calculated with DFT+$U$. (b) Total energy as a function of the distance of the Ho adatom from the graphene sheet calculated with DFT+$U$ (full dots) and the respective fit with a Morse-like potential (line). }
    \label{MAE_vs_d}
\end{figure}

Comparing the different REs, Tm/Gr shows the smallest $|K_4|$ value; Ho and Dy
instead have the same order of magnitude $K_4\sim 1$ meV for $d/d_{0}=0.96$, 
which might reflect a modulation of the charge distribution in the $xy$-plane compared to Tm. 
Overall, a stronger MAE emerges when the rare-earth atom is pressed against the graphene sheet, since the $4f$ and $5d$ electrons of the impurity \textquote{feel} a stronger electrostatic repulsion from the carbon atoms.
Fig.~\ref{MAE_vs_d}(a) shows the energy difference $\Delta E= E_{\parallel}-E_{\perp}$, with $E_{\parallel}$ being the total energy when the magnetization is aligned along the $x$-axis (parallel to the graphene), while $E_{\perp}$ represents the total energy with magnetization along the $z$-axis (perpendicular to the graphene),
as function of $d/d_{0}$. We scan the values of the MAE for values 
ranging in $d/d_{0}=[0.9,1.04]$ using a step of $0.01$. Positive values 
of the $\Delta E$ indicate that an out-of-plane magnetization is favoured compared to an in-plane magnetization. 

For Ho/Gr and Tm/Gr, $\Delta E$ decreases when the adatom 
is compressed towards the graphene from $d/d_{0}=1.04$ to smaller distances, 
going via a minimum and then increasing steeply for high compression of around
$10\%$.
In contrast, Dy undergoes a switch of the favoured magnetization direction, since in case of high compression an out-of-plane is more stable, while the in-plane direction appears lower in energy for larger distances from the graphene. Such a mutable magnetic behavior 
might find interesting applications in engineering magneto-mechanical nano 
devices that rely on pressure-induced magnetization transitions \cite{cenker2022reversible,jiles2003role,hu2020enhanced}.

We show in Fig.~\ref{MAE_vs_d}(b) the evolution of the total energy, reported with respect to the energy minimum (placed at 0 eV), as a function of $d/d_{0}$ considering an 
out-of-plane orientation of the moment for Ho/Gr (we obtain very similar curves for Dy/Tm). The same calculations have been performed also with in-plane magnetization for each system and in total $6$ energy-curves have been obtained and fitted adopting a Morse potential.
The Morse fit is shown in continuous line, and has been performed using Eq.~\eqref{Morse} 
given in Appendix~\ref{Appendix:d}. The fit provides the dissociation energy $D_e$ with respect to the minimum at the equilibrium distance and the width $b$ of the curve.
These values are then 
employed to calculate the force constant $k_e$ at the equilibrium position of 
the oscillator as $k_e=2b^2D_e$, that defines the stiffness against deformation.
The vibrational frequency $\nu$ of the displacement is evaluated following 
Eq.~\eqref{frequency}.
Table~\ref{table5} summarizes the results obtained for the RE/Gr systems: 
$k_e$ and $\nu$ calculated for the two directions of 
the magnetization, namely parallel to the Gr plane ($\parallel$) and perpendicular 
($\perp$). There is a slight dependence of the vibrational frequencies on
the magnetization orientation changes. For Dy, the force constant $k_e$ at 
equilibrium is bigger for an out-of-plane magnetization, meaning the material 
is more resistant against deformation when perpendicularly magnetized, which 
reflects in a slightly higher vibrational frequency of the mode (see Eq.~\eqref{frequency} in Appendix~\ref{Appendix:d}). 
Ho and Tm have a higher $k_e$ and $\nu$ for the in-plane magnetization 
direction. Lastly, among the systems at hand, Tm/Gr displays the lowest $k_e$ 
values, which determines the weakest bonding towards the substrate and makes 
it the most malleable 2D material in the set.

Taking into account the bistability observed in the multiplet spectra of Ho/Gr and Tm/Gr (Fig.~\ref{Ho/Tm_splitting}), there exists a separation between the two ground states with energy gaps of approximately $\Delta E\sim 14$~meV and $\Delta E\sim 18$~meV, respectively. When converting the vibrational frequencies of the rare-earth atoms' modes into vibrational energies, we obtain values ranging from $h\nu=60-70$~meV. This implies that magnetization reversal due to adatom vibrations, which would necessitate $\Delta E=h\nu$, is unlikely.
\begin{table}[H]
\centering
 \hspace{-5mm}
\caption{\label{table5} Elastic force constants $k_e$ ($N/m$) and the respective vibration frequencies $\nu$ ($s^{-1}$) calculated with perpendicular and parallel magnetization for each RE/Gr system.}

\begin{tabular*}{\linewidth}{c@{\extracolsep{\fill}}ccccc}
\midrule[0.4mm]
RE/Gr  & $k_{e\perp}$ &$k_{e\parallel}$ & $\nu_{\perp}\cdot 10^{-13}$ &$\nu_{\parallel}\cdot 10^{-13}$\\
\midrule[0.2mm]
\bfseries{Dy/Gr} &$1155.54$&$1118.13$&$1.879$&$1.848$ \\
\bfseries{Ho/Gr}  &$1029.05$ &$1076.14 $&$1.769$&$1.809$\\
\bfseries{Tm/Gr} &$784.72$&$908.63 $&$1.539$ & $1.656$ \\
\midrule[0.4mm]
\end{tabular*}
\end{table}

\section{Dysprosium on graphene: deviation from Hund's rules}
\label{Dy5}

The calculations presented above are based on the occupation of $4f$-shells following a 
Hund-like ground state. Nonetheless, further analysis shows that for Dy/Gr 
a orbital occupation can be obtained, where one minority spin electron moves from the orbital with quantum number $m_{l}=1$ to the $m_{l}=0$, partially quenching 
the orbital moment to $m^\text{RE}_{l}=4.9\mu_\text{B}$. This indicates that 
for this particular case of Dy/Gr the crystal field effects are strong enough 
to compete with the Hund's exchange. This re-arrangement affects mainly the orbital moment as 
it leaves unaltered $m^\text{RE}_s=4.03\mu_\text{B}$, hence only breaking 
Hund's second rule (maximizes $L$) and leading to a total angular momentum 
$J=7$.
Next, we investigate the behavior of the magnetic anisotropy in this new 
orbital configuration. We will refer to this $4f$ occupation as $J=7$ state 
and to the Hund's rules orbital occupation as $J=8$ state.  Specifically, the energy difference calculated between the two observed magnetic states is~$0.28$~eV in favour of the $J=7$ situation.

\begin{figure}[t!]
\begin{center}
    \hspace{-6mm}
    \includegraphics[scale=0.15]{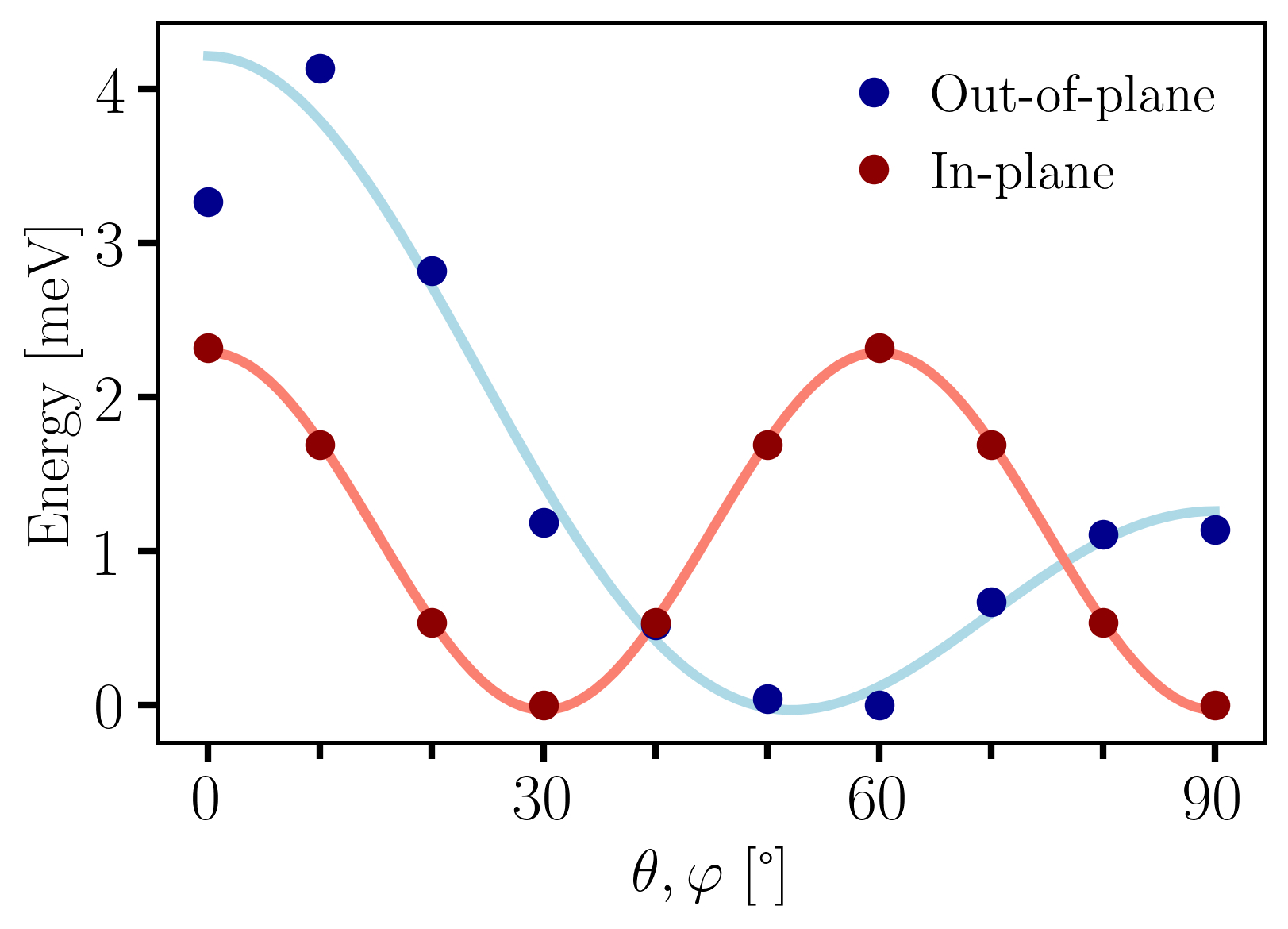}
    \caption{Out-of-plane (blue) and in-plane (red) magnetic anisotropy energy curves for Dy/Gr with Dy with orbital momentum $m_{l}^\mathrm{RE}=5$ $\mu_\text{B}$ . The blue dots indicate the DFT+$U$ energies, while the line corresponds to the fitting.}
    \label{MAE_Dy5}
\end{center}
\end{figure}

\begin{figure*}[t!]
\centering
    \includegraphics[scale=0.15]{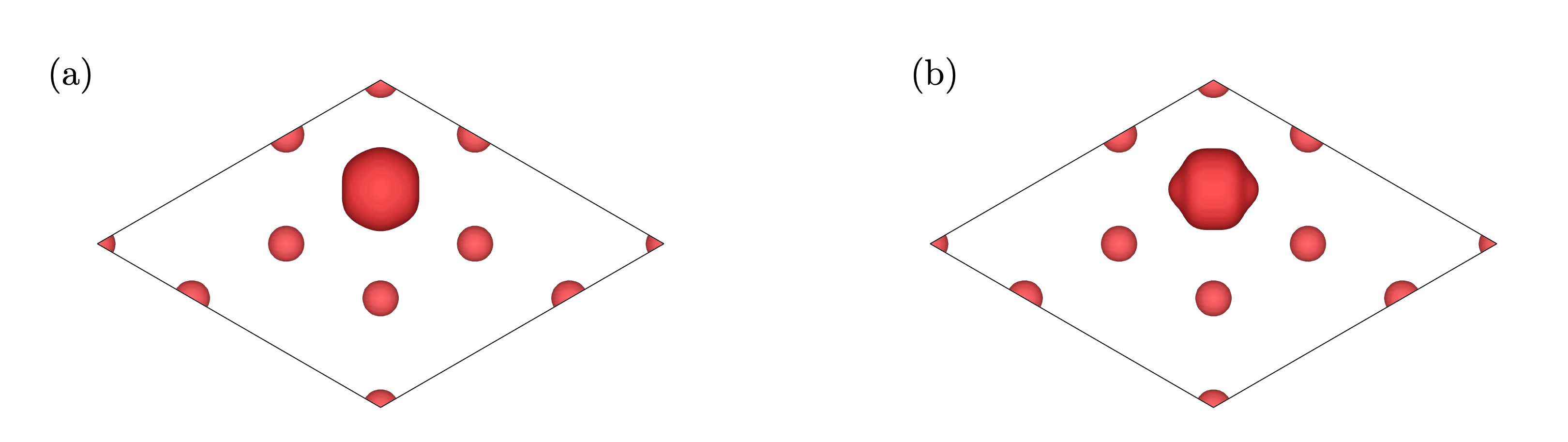}
    \caption{Total charge density of the RE and Gr atoms of the spin-down channel of Dy/Gr with in-plane magnetization for the two different orbital moments: (a) $m_l=6$ $\mu_\text{B}$  ($J=8$) (b) $m_l=5$ $\mu_\text{B}$ ($J=7$). For atom positions compare with Fig.~\ref{Gr-sites} with the Dy atom in the H-site.}
    \label{Dy_spindensity}
\end{figure*}

Given the close link between the orbital moment and the MAE, we expect a deviation from the 
angular dependence determined for the $J=8$ state shown in Fig.~\ref{MAE_curves}. 
This distinct angular form is attributed to the new shape of the $4f$ charge 
cloud in the $J=7$ state with respect to the $J=8$ state as depicted in 
Fig.~\ref{Dy_spindensity}. Indeed, the angular dependence of the MAE curve is driven by the interplay of the charge cloud's geometry and the CF symmetry: a rotation of the magnetization corresponds to a rotation of the anisotropic charge distribution through SOC interaction and can thus lead to stronger/weaker electrostatic repulsion if the charge cloud lies closer/farther away from the point charges defining the CF.
Fig.~\ref{MAE_Dy5} shows the respective out-of-plane and in-plane MAE curves,
where the dots (full lines) represent the DFT+$U$ data (fits). 
The fitted MAE 
coefficients $K_i$ using Eq.~\eqref{MAE} are given in Table~\ref{table8}, 
the out-of-plane anisotropy has a minimum at a canted angle $\theta\simeq51.82^{\circ}$, 
while the basal-anisotropy favours an angle $\varphi=30^{\circ}\pm \frac{\pi}{3}$. 

For the out-of-plane case, the shape of the MAE curve is drastically different
when comparing the $J=7$ state (Fig.~\ref{MAE_Dy5}) with the $J=8$ state 
(Fig.~\ref{MAE_curves}a). Here, instead of an energy barrier between 
the perpendicular magnetization ($\theta=0^{\circ}$) and the in-plane 
magnetization ($\theta=\frac{\pi}{2}^{\circ}$), there is an energy valley.
These trends are reflected in the $K_i$ coefficients, which have opposite 
signs when comparing the two orbital configurations 
(see Table~\ref{table8}).
\begin{table}[b!]
\centering
\caption{\label{table8} Magnetic anisotropy constants $K_i$ for the $J=7$ and $J=8$ states in Dy/Gr. Results are in meV.}
\begin{tabular*}{\linewidth}{@{\extracolsep{\fill}}
    l*{5}{S[table-format=-1.3]}}
          % {l *{5}{S[table-format=-1.4]}}%{@{}lllll}
\midrule[0.4mm]
\mc{$State$} &\mc{$K_{1}$} & \mc{$K_{2}$}& \mc{$K_{3}$} & \mc{$K_{4}$}\\
\midrule[0.2mm]
$J$=7 &-14.29& 13.10&-1.76& 1.16\\ 
$J$=8 &15.36&-18.92& 1.54&-0.44 \\ 
\midrule[0.4mm]
\end{tabular*}
\end{table}
Interestingly, the in-plane MAE shown in red in Fig.~\ref{MAE_Dy5} 
is boosted and reaches values of the same order of magnitude as its 
out-of-plane counterpart. It is the largest in-plane MAE observed for
all the systems at hand and this enhancement is reflected in the $K_{4}$
coefficient which causes a global minima ($\frac{\partial E_{an}}{\partial \theta}|_{\varphi=30^{\circ}}$) at $(\theta,\varphi)=(57.93^{\circ},\varphi=30^{\circ}\pm\frac{\pi}{3})$.
We note that this energy minimum 
does not coincide with minimum defined by the purely out-of-plane MAE ($\frac{\partial E_{an}}{\partial \theta}|_{\varphi=0^{\circ}}$) 
located at $\theta=51.82^{\circ}$. Qualitatively, this large $K_4$ can be understood due to the shape of the $4f$-charge density in Fig.~\ref{Dy_spindensity} that shows the spin-down charge density computed for the in-plane magnetization $\parallel x$-axis. Indeed, for the $J=7$ state there 
is a larger spin-down density in 
the $xy$-plane ($\parallel$ to the substrate) in contrast to the $J=8$ state, for which the charge density displays smaller poles in the $xy$-plane. 
These findings demonstrate the necessity of a careful assessment of the $4f$-orbital occupation when dealing with rare-earth based systems in low dimensions. This is also important when wanting to adopt the before proposed reverse-engineering method to determine the CFP, since it holds only in the context of a Hund's rule occupation, when the interelectronic repulsion and SOC effects dominate and the crystal field can be treated as a perturbation. Further advanced methods such as the Hubbard-I approximation~\cite{shick2009HIA} will be considered in the future along with a comparison to experimental data.

%\SB{\section{Reflection on choice of unit cell size}
%Here we could put the discussion satisfying Referee 2 + 3 adding a bit a longer text along the line of my reply to Refere 2 point 1.}

%\SB{Since the properties of RE atoms on graphene or related substrate  are crucially determined by the 4f electrons and the 4f electrons are localized we work with a rather samll  root 3 times root e unit cell to investigte the properties of the properties...., Refere to the experiments mention above ...  }

\section{Conclusion}
In this work, we investigated the electronic structure of a selected 
subset of $4f$ adatoms (Dy, Ho and Tm) deposited on graphene and treated
the localized $4f$-electrons using the DFT+$U$ approach. The RE/Gr complexes
display a metallic behavior due to $n$-doping originating from the rare-earth's
$d$-orbitals. 
In all the analyzed RE/Gr systems, the rare-earth atoms adopt a divalent configuration \rm{RE}$^{2+}$ in which the orbital occupation of Ho and Tm is consistent with a maximal orbital moment, while Dy displays a lower energy for a $J=7$ configuration instead of $J=8$ (Hund's rules). This deviation originates from the competition between the crystal field and intra-atomic exchange. 
The $f$-states maintain 
a localized behavior and carry a high orbital moment which results in a 
consequently large single-ion magnetic anisotropy. The self-consistent total energy calculations show barriers of several meV upon variation of the magnetization
direction.

From the MAE curves, we extracted the magnetic anisotropy constants which are then reverse-engineered to crystal field parameters (CFP) in the Steven convention. The obtained CFP are then adopted in the diagonalization of the CF Hamiltonian matrix in the $C_{6v}$ symmetry to calculate the multiplet structure of each system. Half-integer spin systems (Ho/Gr and Tm/Gr) do not present tunnel-split doublets at $  \braket{J_z}=0$ and are thus protected against quantum tunneling of magnetization via such states, while in Dy/Gr in the $J=8$ state, we find that the high-order crystal field parameter $C_6^6$ generates potential states at quenched $\braket{J_z}$ that might reduce the energy barrier for spin reversal. Nevertheless, in the latter case we find a single magnetic ground state at $\braket{J_z}=0$ with no possibility of reversal. Further studies to determine systems with high magnetic anisotropy energies and protection against magnetization reversal might involve the analysis of the effect of the symmetry of different substrates on the multiplet splittings as well as the impact of the chemical composition of the crystal field, \textit{i.e.}\ inducing stronger SOC and/or orbital hybridizations to the adsorbed RE atom.

The analysis of the magnetic anisotropy is then further extended to inspect magneto-elastic effects. The application of a perpendicular strain compressing the adatom towards the graphene enhances the magnetic anisotropy, thus providing another mechanism to amplify the magnetic stability of these
$4f$-adatoms. Increasing the adatom-substrate distance leads to a decoupling from the substrate, driving the atom to a quasi-isolated state, ultimately reducing the magnetic anisotropy. A deviation from this behavior has been observed for the in-plane anisotropy constant $K_4$ in the case of Tm/Gr, where a non-monotonous trend is found as a function of the strain. 
For the particular case of Dy/Gr, the mechanical strain induces a change in the sign of the energy difference $\Delta E= E_{\parallel}-E_{\perp}$ in the $J=8$ magnetic state, indicating the possibility to tailor the favoured magnetization direction by application of an external stress.
Finally, our first principles investigation emphasizes the necessity of a detailed analysis of the orbital occupations of these $4f$-compounds, as this leads qualitatively and quantitatively different magnetic anisotropies.

\section*{Acknowledgments} 
J.P.C.\ acknowledges Dr.\ Eduardo Mendive Tapia for the inspiring discussions concerning magnetoelastic phenomena; and Henning Janßen for advices on FLEUR calculations.
The project is funded by the Deutsche Forschungsgemeinschaft (DFG) through CRC
1238, Control and Dynamics of Quantum Materials: Spin
orbit coupling, correlations, and topology (Project No.\ C01). 
This research is also supported by the FLAG-ERA grant SOgraphMEM, Project PCI2019-111867-2. 
We acknowledge computing resources granted by RWTH Aachen University
under Project No.\ jara0219.

\appendix
\section{}
\setcounter{section}{1}

\label{appendix:a}
The Steven's operators $\hat{O}_{l}^m$ for a $C_{6v}$ crystal field write as

\begin{equation}
    \begin{split}
        \hat{O}_2^0=&3\hat{J}_z^2-X\\
        \hat{O}_4^0=&35\hat{J}_z^4-(30X-25)\hat{J}_z^2+3X^2-6X\\
        \hat{O}_6^0=&231\hat{J}_z^6-(315X-735)\hat{J}_z^4\\
        &+(105X^2-525X+294)\hat{J}_z^2\\
        &-5X^3+40X^2-60X \\
        \hat{O_6^6}=&\frac{1}{2}\left[\hat{J}_+^6+\hat{J}_-^6 \right]
    \end{split}
    \label{StevensOp}
\end{equation}
where $X=J(J+1)$, $\hat{J}_+=\hat{J}_x+i\hat{J}_y$, $\hat{J}_-=\hat{J}_x-i\hat{J}_y$.
The selection rules for non-zero elements of the $lm$-expansion are dictated by the lattice symmetry. For the $4f$-shell in a $C_{6v}$ symmetry, the expansion of $\hat{H}_{\text{CF}}$ is defined by the quantum numbers $l=\{0,2,4,6\}$ and $m=0,6$.
\section{}
\label{Appendix:b}
In the following, the angular functions appearing in Eq.~\eqref{CF_energ} are defined. 
\begin{equation}
\begin{split}
   F_2^0(\theta)&=-3\sin^2\theta \\
   F_4^0(\theta)&=35\sin^4\theta-40\sin^2\theta \\
   F_6^0(\theta)&=-231\sin^6\theta+378\sin^4\theta-168\sin^2\theta \\
   F_6^6(\theta)&=\sin^6\theta \\
   G_0(\varphi)&=1 \\
   G_6(\varphi)&=\cos\varphi
\end{split}
\label{angular_func}
\end{equation}

The $F_l^m$ functions define pure out-of-plane rotation and the $G_m$ functions are related to in-plane rotations.

\section{}
\label{Appendix:c}

\begin{figure}[t!]
    \includegraphics[scale=0.55]{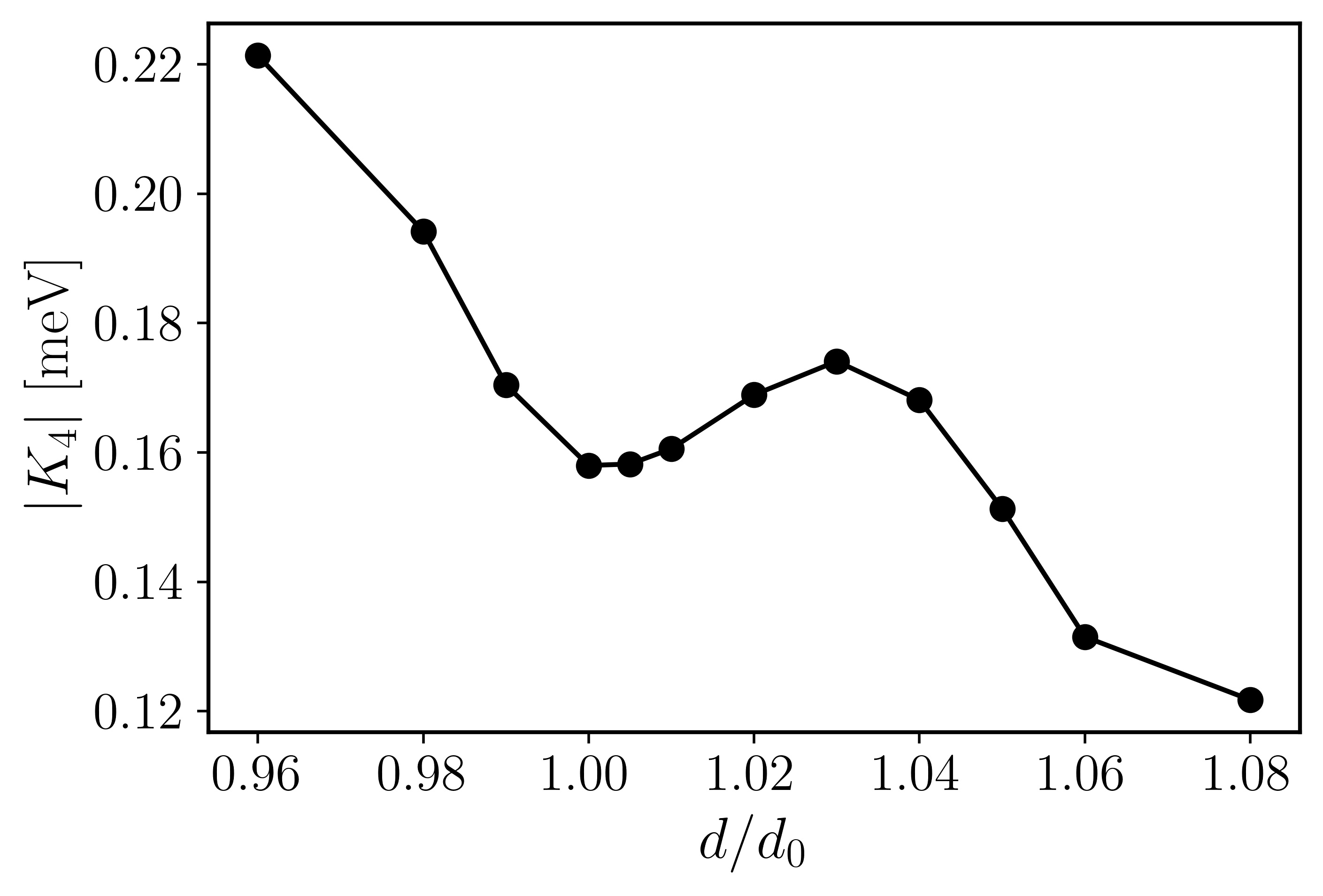}
    \caption{$|K_4|$ of Tm/Gr for distances ranging from $d=0.96$ to $d=1.08$ from the graphene sheet \textit{i.e.} from $-4\%$ to $+8\%$ of  perpendicular strain. }
    \label{Tm_K4}
\end{figure}

In Fig.~\ref{Tm_K4} the modulus of the magnetic anisotropy constant $K_4$ for Tm/Gr is plotted against different perpendicular strains $d/d_0$ from the graphene monolayer. Overall, the modulus $|K_4|$ displays a non-linear behavior first increasing from $d/d_0=1.08$ until a maximum is reached at  $d/d_0=1.03$, then the module shows a decreasing trend until $d/d_0=1.0$, after which $|K_4|$ grows again for small distances.

\section{}
\label{Appendix:d}

The potential energy function adopted in the fitting of the data in Fig.~\ref{MAE_vs_d}(b) of section~\ref{magnetoel_section} reads as \cite{Morse1,Morse2}
\begin{equation}
    V(r)=D_e\left(1-e^{-b(d-d_0)}\right)^2\, ,
    \label{Morse}
\end{equation}
where $D_e$ corresponds to the depth of the potential with respect to the dissociation energy, $d$ is the distance between RE and Gr, and $d_0$ is the equilibrium distance. $b$ determines the width of the potential well.
The frequency of the vibrational modes in the RE/Gr complexes is evaluated as in a diatomic-like system
\begin{equation}
    \nu=\frac{1}{2\pi}\sqrt{\frac{k_e}{\mu}}
    \label{frequency}
\end{equation}
being $k_e$ the force constant of the RE-Gr interaction and  $\mu$ the reduced mass $\mu=\frac{m_{\rm{Gr}}m_{\rm{RE}}}{m_{\rm{Gr}}+m_{\rm{RE}}}$ of the RE/Gr complex, with $m_{\rm{RE}}$ the atomic mass of the RE atom and $m_{\rm{Gr}}$ the mass of graphene in the considered simulation cell involving $6$ carbon atoms, equal to $72.066$ amu. 
\newpage
\section{}
\label{Appendix:e}
For the determination of the multiplet structures we follow the tables in Ref.~\cite{hutchings1964point} for the calculation of the matrix elements of the crystal field Hamiltonian in Eq.~\ref{CF}. In the following the CF matrix for Tm is provided as an example.
From the DFT+$U$ calculations it is understood that in Tm/Gr the $4f$ electrons closely follow Hund's rules, providing a total angular momentum $J=\frac{7}{2}$, and hence $J_z$ taking values $\{-\frac{7}{2},-\frac{5}{2},-\frac{3}{2},-\frac{1}{2},\frac{1}{2},\frac{3}{2},\frac{5}{2},\frac{7}{2}\}$.
The respective $8\times8$ crystal field matrix is given by 
\begin{equation}
\begin{pmatrix}
A & 0 & 0 & 0 & 0 & 0 & E &0 \\
0& B & 0&0& 0&0& 0&E \\
0& 0&C & 0& 0&  0& 0& 0\\
0&0&  0& D& 0&  0& 0& 0\\
0&0& 0& 0& D&  0& 0& 0\\
0&0&0&0& 0& C & 0& 0 \\
E& 0&  0& 0& 0&  0&B & 0\\
0&E& 0&  0& 0&  0& 0&A\\ 
\end{pmatrix} 
\label{CF_matrix}
\end{equation}
and the matrix elements  $\braket{J_{z}=i|\hat{H}_{\text{CF}}|J_{z}=j}$ are defined as 
\begin{widetext}
\begin{equation}
\begin{split}
 A&=\braket{\frac{7}{2}|\hat{H}_{\text{CF}}|\frac{7}{2}}=\braket{-\frac{7}{2}|\hat{H}_{\text{CF}}|-\frac{7}{2}}=21C_2^0+420C_4^0+1260C_6^0 \\
 B& =\braket{\frac{5}{2}|\hat{H}_{\text{CF}}|\frac{5}{2}}=\braket{-\frac{5}{2}|\hat{H}_{\text{CF}}|-\frac{5}{2}}=3C_2^0-780C_4^0-6300C_6^0\\
 C&=\braket{\frac{3}{2}|\hat{H}_{\text{CF}}|\frac{3}{2}}=\braket{-\frac{3}{2}|\hat{H}_{\text{CF}}|-\frac{3}{2}}=-9C_2^0-180C_4^0+11340C_6^0\\
 D&=\braket{\frac{1}{2}|\hat{H}_{\text{CF}}|\frac{1}{2}}=\braket{-\frac{1}{2}|\hat{H}_{\text{CF}}|-\frac{1}{2}}=-15C_2^0+540C_4^0-6300C_6^0\\
 E&=\braket{\frac{7}{2}|\hat{H}_{\text{CF}}|-\frac{5}{2}}=\braket{-\frac{7}{2}|\hat{H}_{\text{CF}}|\frac{5}{2}}=\braket{\frac{5}{2}|\hat{H}_{\text{CF}}|-\frac{7}{2}}=\braket{-\frac{5}{2}|\hat{H}_{\text{CF}}|\frac{7}{2}}=360\sqrt{7}C_6^6
   \end{split}
   \label{matrixelements}
\end{equation}
\end{widetext}
where the $C_l^m$ are the CF parameters. The exact values in front of each CFP correspond to the matrix element of the respective Steven's operator between the same states ($i=j$) for the diagonal terms, for example $\braket{\frac{7}{2}|\hat{O}_2^0|\frac{7}{2}}=21$, while the only non-zero terms of $\braket{J_{z}=i|\hat{O}_6^6|J_{z}=j}$ with $i\neq j$ are those between $J_z$ states differing by $6$.
Hence, the matrix is symmetric and presents non-zero off-diagonal terms for $J_z$ state differing by $\Delta J_z=\pm 6$, which are mixed by the $C_6^6$ operator. The CF matrix is set up by inserting the $C_l^m$ values obtained via DFT+$U$ calculations and diagonalized, leading to $8$ eigenvectors in the case of Tm.

Concerning the degeneracy of the states, it can be worked out following the orthogonality theorem to determine how the energy levels of an isolated spherically symmetric ($K_h$) Tm atom split into a sum of irreducible representations (IR) of the $C_{6v}$ point-group.
\begin{equation}
    \sum_{\nu=1}^{n}\chi^{(\alpha)}(g_{\nu})[\chi^{(\beta)}(g_{\nu})]^{*}=n\delta_{\alpha\beta}    \label{Orthogonality}
\end{equation}
Here, $\chi^{(\alpha)}$ and $\chi^{(\beta)}$ are the characters of the two compared representations $\alpha$ and $\beta$, and the index $\nu$ sums over the number $n$ of symmetry operations, which is $n=24$ for the $C_{6v}$ case. The character table of the two space groups is shown in Table~\ref{tab:1E}.
\begin{table}[b]
\centering
\begin{tabular}{c|rrrrrrrrr}
\hline
 $C_{6v}$& $E$ & $\bar{E}$ & $C_{2}$ &$2C_{3}$  & $\bar{2C_{3}}$ & $2C_{6}$ & $\bar{2C_{6}}$ & $3\sigma_{d}$ & $3\sigma_{v}$ \\
 & & & $\bar{C_{2}}$ &  & &  &  & $3\bar{\sigma_{d}}$&$3\bar{\sigma_{v}}$\\
\hline
$\Gamma_{1}$ & 1 & 1 & 1 & 1 & 1 & 1 & 1 & 1 &1\\
$\Gamma_{2}$ & 1 & 1 & 1 & 1 & 1 & 1 & 1 & $-$1 & $-$1 \\
$\Gamma_{3}$ & 1 & 1 & $-$1 & 1 & 1 & $-$1 &$-$1 & 1 &$-$1 \\
$\Gamma_{4}$ & 1 & 1 & $-$1 & 1 & 1 & $-$1 & $-$1 & 1 &1 \\
\hline
$\Gamma_{5}$ & 2 & 2 & $-$2 & $-$1 & $-$1 & 1 & 1 & 0 &0 \\
$\Gamma_{6}$ & 2 & 2 & 2 & $-$1 & $-$1 & $-$1&  $-$1 & 0 & 0\\
$\Gamma_{7}$ & 2 & $-$2 & 0 & 1 & $-$1 & $\sqrt{3}$ & $-\sqrt{3}$ & 0 &0\\
$\Gamma_{8}$ & 2 & $-$2 & 0 & 1 & $-$1 & $-\sqrt{3}$ & $\sqrt{3}$ & 0 &0\\
$\Gamma_{9}$ & 2 & $-$2 & 0 & $-$2 & 2 & 0 & 0& 0 &0\\
\hline \hline
Tm $K_h$ & 8 & $-$8 & 0 & 1 & $-$1 & $-\sqrt{3}$ & $\sqrt{3}$ & 0 & 0\\
\hline
\end{tabular}
\caption{Character table of the $C_{6v}$ symmetry group and of the rotational invariant Tm atom with total angular momentum $J=\frac{7}{2}$.}
\label{tab:1E}
\end{table}
\begin{figure*}[t!]
\centering
    \includegraphics[scale=0.14]{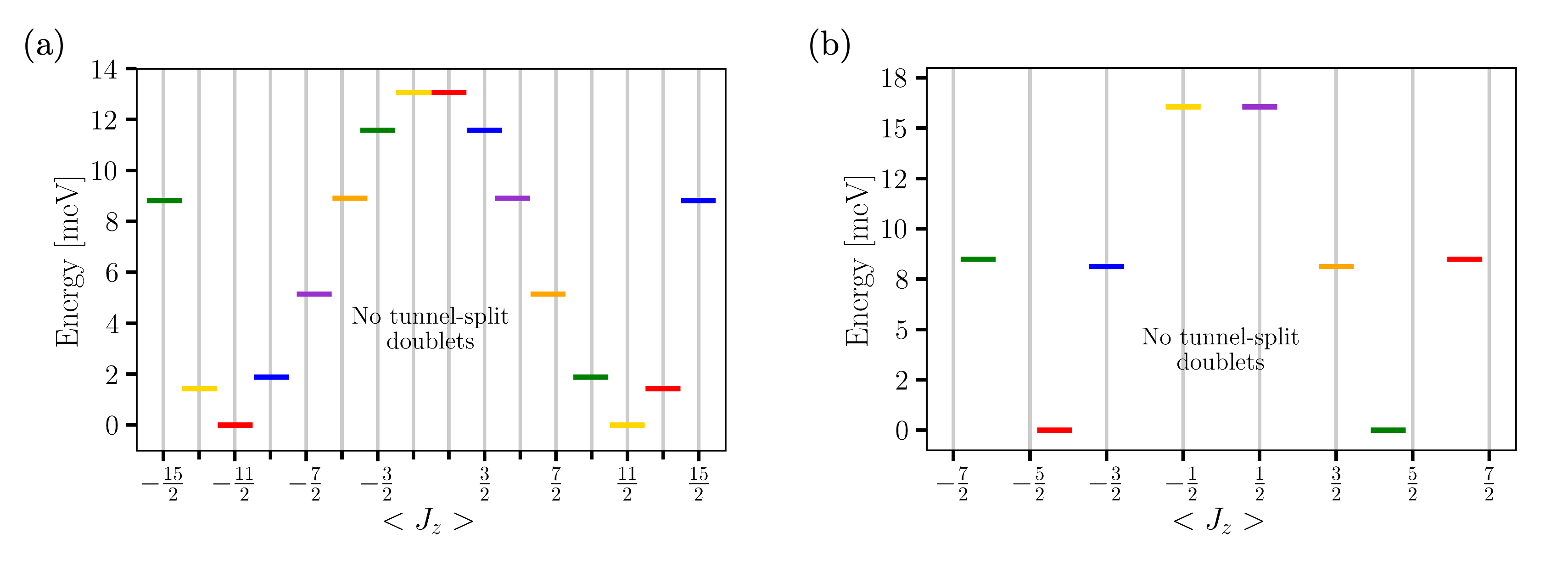}
    \caption{Multiplet splitting of (a) Ho/Gr and (b) Tm/Gr adopting the CFP values obtained from reverse-engineering via the magnetic anisotropy constants. States in the same color correspond to linear combinations of $\ket{J_z}$ differing by $\Delta J_z=\pm6,\pm12$.}
    \label{Ho/Tm_splitting}
\end{figure*}

To obtain the splitting, one multiplies the character of each operation of the $K_h$ group with the respective character of the $C_{6v}$ group and sums this value over all symmetry operations for each IR in the $C_{6v}$ group separately. The characters are orthogonal which means that either this sum gives 0 and thus the respective IR of the $C_{6v}$ group is not included or the sum is an integer, $n$, which tells how many times the IR is included. In the case of Tm/Gr Eq.~\eqref{Orthogonality} gives rise to a splitting of $4$ sets of double degenerate state, two sets of which belonging to the same IR, namely $\Gamma_8$.

\begin{equation}
      K_{h}^{7/2}=\Gamma_{7}+2\Gamma_{8}+\Gamma_{9}
\end{equation}

The multiplets for Tm/Gr and Ho/Gr are shown in Fig.~\ref{Ho/Tm_splitting}.

\bibliography{main.bib}

\begin{thebibliography}{65}%
\makeatletter
\providecommand \@ifxundefined [1]{%
 \@ifx{#1\undefined}
}%
\providecommand \@ifnum [1]{%
 \ifnum #1\expandafter \@firstoftwo
 \else \expandafter \@secondoftwo
 \fi
}%
\providecommand \@ifx [1]{%
 \ifx #1\expandafter \@firstoftwo
 \else \expandafter \@secondoftwo
 \fi
}%
\providecommand \natexlab [1]{#1}%
\providecommand \enquote  [1]{``#1''}%
\providecommand \bibnamefont  [1]{#1}%
\providecommand \bibfnamefont [1]{#1}%
\providecommand \citenamefont [1]{#1}%
\providecommand \href@noop [0]{\@secondoftwo}%
\providecommand \href [0]{\begingroup \@sanitize@url \@href}%
\providecommand \@href[1]{\@@startlink{#1}\@@href}%
\providecommand \@@href[1]{\endgroup#1\@@endlink}%
\providecommand \@sanitize@url [0]{\catcode `\\12\catcode `\$12\catcode
  `\&12\catcode `\#12\catcode `\^12\catcode `\_12\catcode `\%12\relax}%
\providecommand \@@startlink[1]{}%
\providecommand \@@endlink[0]{}%
\providecommand \url  [0]{\begingroup\@sanitize@url \@url }%
\providecommand \@url [1]{\endgroup\@href {#1}{\urlprefix }}%
\providecommand \urlprefix  [0]{URL }%
\providecommand \Eprint [0]{\href }%
\providecommand \doibase [0]{https://doi.org/}%
\providecommand \selectlanguage [0]{\@gobble}%
\providecommand \bibinfo  [0]{\@secondoftwo}%
\providecommand \bibfield  [0]{\@secondoftwo}%
\providecommand \translation [1]{[#1]}%
\providecommand \BibitemOpen [0]{}%
\providecommand \bibitemStop [0]{}%
\providecommand \bibitemNoStop [0]{.\EOS\space}%
\providecommand \EOS [0]{\spacefactor3000\relax}%
\providecommand \BibitemShut  [1]{\csname bibitem#1\endcsname}%
\let\auto@bib@innerbib\@empty
%</preamble>
\bibitem [{\citenamefont {Guo}\ \emph {et~al.}(2021)\citenamefont {Guo},
  \citenamefont {Hu}, \citenamefont {Liu},\ and\ \citenamefont
  {Tian}}]{guo2021stacking}%
  \BibitemOpen
  \bibfield  {author} {\bibinfo {author} {\bibfnamefont {H.-W.}\ \bibnamefont
  {Guo}}, \bibinfo {author} {\bibfnamefont {Z.}~\bibnamefont {Hu}}, \bibinfo
  {author} {\bibfnamefont {Z.-B.}\ \bibnamefont {Liu}},\ and\ \bibinfo {author}
  {\bibfnamefont {J.-G.}\ \bibnamefont {Tian}},\ }\href
  {https://doi.org/10.1002/adfm.202007810} {\bibfield  {journal} {\bibinfo
  {journal} {Advanced Functional Materials}\ }\textbf {\bibinfo {volume}
  {31}},\ \bibinfo {pages} {2007810} (\bibinfo {year} {2021})}\BibitemShut
  {NoStop}%
\bibitem [{\citenamefont {Miró}\ \emph {et~al.}(2014)\citenamefont {Miró},
  \citenamefont {Audiffred},\ and\ \citenamefont {Heine}}]{Miro2014}%
  \BibitemOpen
  \bibfield  {author} {\bibinfo {author} {\bibfnamefont {P.}~\bibnamefont
  {Miró}}, \bibinfo {author} {\bibfnamefont {M.}~\bibnamefont {Audiffred}},\
  and\ \bibinfo {author} {\bibfnamefont {T.}~\bibnamefont {Heine}},\ }\bibfield
   {title} {\bibinfo {title} {An atlas of two-dimensional materials},\ }\href
  {https://doi.org/10.1039/C4CS00102H} {\bibfield  {journal} {\bibinfo
  {journal} {Chem. Soc. Rev.}\ }\textbf {\bibinfo {volume} {43}},\ \bibinfo
  {pages} {6537} (\bibinfo {year} {2014})}\BibitemShut {NoStop}%
\bibitem [{\citenamefont {Manzeli}\ \emph {et~al.}(2017)\citenamefont
  {Manzeli}, \citenamefont {Ovchinnikov}, \citenamefont {Pasquier},
  \citenamefont {Yazyev},\ and\ \citenamefont {Kis}}]{manzeli20172d}%
  \BibitemOpen
  \bibfield  {author} {\bibinfo {author} {\bibfnamefont {S.}~\bibnamefont
  {Manzeli}}, \bibinfo {author} {\bibfnamefont {D.}~\bibnamefont
  {Ovchinnikov}}, \bibinfo {author} {\bibfnamefont {D.}~\bibnamefont
  {Pasquier}}, \bibinfo {author} {\bibfnamefont {O.~V.}\ \bibnamefont
  {Yazyev}},\ and\ \bibinfo {author} {\bibfnamefont {A.}~\bibnamefont {Kis}},\
  }\href {https://doi.org/10.1038/natrevmats.2017.33} {\bibfield  {journal}
  {\bibinfo  {journal} {Nature Reviews Materials}\ }\textbf {\bibinfo {volume}
  {2}},\ \bibinfo {pages} {1} (\bibinfo {year} {2017})}\BibitemShut {NoStop}%
\bibitem [{\citenamefont {Novoselov}\ \emph {et~al.}(2016)\citenamefont
  {Novoselov}, \citenamefont {Mishchenko}, \citenamefont {Carvalho},\ and\
  \citenamefont {Castro~Neto}}]{novoselov2016}%
  \BibitemOpen
  \bibfield  {author} {\bibinfo {author} {\bibfnamefont {K.}~\bibnamefont
  {Novoselov}}, \bibinfo {author} {\bibfnamefont {A.}~\bibnamefont
  {Mishchenko}}, \bibinfo {author} {\bibfnamefont {A.}~\bibnamefont
  {Carvalho}},\ and\ \bibinfo {author} {\bibfnamefont {A.}~\bibnamefont
  {Castro~Neto}},\ }\href {https://doi.org/10.1126/science.aac9439} {\bibfield
  {journal} {\bibinfo  {journal} {Science}\ }\textbf {\bibinfo {volume}
  {353}},\ \bibinfo {pages} {aac9439} (\bibinfo {year} {2016})}\BibitemShut
  {NoStop}%
\bibitem [{\citenamefont {Liu}\ \emph {et~al.}(2016)\citenamefont {Liu},
  \citenamefont {Weiss}, \citenamefont {Duan}, \citenamefont {Cheng},
  \citenamefont {Huang},\ and\ \citenamefont {Duan}}]{liu2016van}%
  \BibitemOpen
  \bibfield  {author} {\bibinfo {author} {\bibfnamefont {Y.}~\bibnamefont
  {Liu}}, \bibinfo {author} {\bibfnamefont {N.~O.}\ \bibnamefont {Weiss}},
  \bibinfo {author} {\bibfnamefont {X.}~\bibnamefont {Duan}}, \bibinfo {author}
  {\bibfnamefont {H.-C.}\ \bibnamefont {Cheng}}, \bibinfo {author}
  {\bibfnamefont {Y.}~\bibnamefont {Huang}},\ and\ \bibinfo {author}
  {\bibfnamefont {X.}~\bibnamefont {Duan}},\ }\href
  {https://doi.org/10.1038/natrevmats.2016.42} {\bibfield  {journal} {\bibinfo
  {journal} {Nature Reviews Materials}\ }\textbf {\bibinfo {volume} {1}},\
  \bibinfo {pages} {1} (\bibinfo {year} {2016})}\BibitemShut {NoStop}%
\bibitem [{\citenamefont {Geim}\ and\ \citenamefont
  {Grigorieva}(2013)}]{geim2013van}%
  \BibitemOpen
  \bibfield  {author} {\bibinfo {author} {\bibfnamefont {A.~K.}\ \bibnamefont
  {Geim}}\ and\ \bibinfo {author} {\bibfnamefont {I.~V.}\ \bibnamefont
  {Grigorieva}},\ }\href {https://doi.org/10.1038/nature12385} {\bibfield
  {journal} {\bibinfo  {journal} {Nature}\ }\textbf {\bibinfo {volume} {499}},\
  \bibinfo {pages} {419} (\bibinfo {year} {2013})}\BibitemShut {NoStop}%
\bibitem [{\citenamefont {Ahn}(2020)}]{ahn20202d}%
  \BibitemOpen
  \bibfield  {author} {\bibinfo {author} {\bibfnamefont {E.~C.}\ \bibnamefont
  {Ahn}},\ }\href {https://doi.org/10.1038/s41699-020-0152-0} {\bibfield
  {journal} {\bibinfo  {journal} {npj 2D Materials and Applications}\ }\textbf
  {\bibinfo {volume} {4}},\ \bibinfo {pages} {1} (\bibinfo {year}
  {2020})}\BibitemShut {NoStop}%
\bibitem [{\citenamefont {Jensen}\ and\ \citenamefont
  {Mackintosh}(1991)}]{jensen1991rare}%
  \BibitemOpen
  \bibfield  {author} {\bibinfo {author} {\bibfnamefont {J.}~\bibnamefont
  {Jensen}}\ and\ \bibinfo {author} {\bibfnamefont {A.~R.}\ \bibnamefont
  {Mackintosh}},\ }\href@noop {} {\emph {\bibinfo {title} {Rare-earth
  magnetism}}}\ (\bibinfo  {publisher} {Clarendon Press Oxford},\ \bibinfo
  {year} {1991})\BibitemShut {NoStop}%
\bibitem [{\citenamefont {Schuh}\ \emph {et~al.}(2012)\citenamefont {Schuh},
  \citenamefont {Miyamachi}, \citenamefont {Gerstl}, \citenamefont {Geilhufe},
  \citenamefont {Hoffmann}, \citenamefont {Ostanin}, \citenamefont {Hergert},
  \citenamefont {Ernst},\ and\ \citenamefont {Wulfhekel}}]{schuh2012magnetic}%
  \BibitemOpen
  \bibfield  {author} {\bibinfo {author} {\bibfnamefont {T.}~\bibnamefont
  {Schuh}}, \bibinfo {author} {\bibfnamefont {T.}~\bibnamefont {Miyamachi}},
  \bibinfo {author} {\bibfnamefont {S.}~\bibnamefont {Gerstl}}, \bibinfo
  {author} {\bibfnamefont {M.}~\bibnamefont {Geilhufe}}, \bibinfo {author}
  {\bibfnamefont {M.}~\bibnamefont {Hoffmann}}, \bibinfo {author}
  {\bibfnamefont {S.}~\bibnamefont {Ostanin}}, \bibinfo {author} {\bibfnamefont
  {W.}~\bibnamefont {Hergert}}, \bibinfo {author} {\bibfnamefont
  {A.}~\bibnamefont {Ernst}},\ and\ \bibinfo {author} {\bibfnamefont
  {W.}~\bibnamefont {Wulfhekel}},\ }\href {https://doi.org/10.1021/nl302250n}
  {\bibfield  {journal} {\bibinfo  {journal} {Nano letters}\ }\textbf {\bibinfo
  {volume} {12}},\ \bibinfo {pages} {4805} (\bibinfo {year}
  {2012})}\BibitemShut {NoStop}%
\bibitem [{\citenamefont {Donati}\ \emph {et~al.}(2016)\citenamefont {Donati},
  \citenamefont {Rusponi}, \citenamefont {Stepanow}, \citenamefont
  {Wäckerlin}, \citenamefont {Singha}, \citenamefont {Persichetti},
  \citenamefont {Baltic}, \citenamefont {Diller}, \citenamefont {Patthey},
  \citenamefont {Fernandes}, \citenamefont {Dreiser}, \citenamefont {\u{Z}.
  \u{S}ljivančanin}, \citenamefont {Kummer}, \citenamefont {Nistor},
  \citenamefont {Gambardella},\ and\ \citenamefont {Brune}}]{Gambardella}%
  \BibitemOpen
  \bibfield  {author} {\bibinfo {author} {\bibfnamefont {F.}~\bibnamefont
  {Donati}}, \bibinfo {author} {\bibfnamefont {S.}~\bibnamefont {Rusponi}},
  \bibinfo {author} {\bibfnamefont {S.}~\bibnamefont {Stepanow}}, \bibinfo
  {author} {\bibfnamefont {C.}~\bibnamefont {Wäckerlin}}, \bibinfo {author}
  {\bibfnamefont {A.}~\bibnamefont {Singha}}, \bibinfo {author} {\bibfnamefont
  {L.}~\bibnamefont {Persichetti}}, \bibinfo {author} {\bibfnamefont
  {R.}~\bibnamefont {Baltic}}, \bibinfo {author} {\bibfnamefont
  {K.}~\bibnamefont {Diller}}, \bibinfo {author} {\bibfnamefont
  {F.}~\bibnamefont {Patthey}}, \bibinfo {author} {\bibfnamefont
  {E.}~\bibnamefont {Fernandes}}, \bibinfo {author} {\bibfnamefont
  {J.}~\bibnamefont {Dreiser}}, \bibinfo {author} {\bibnamefont {\u{Z}.
  \u{S}ljivančanin}}, \bibinfo {author} {\bibfnamefont {K.}~\bibnamefont
  {Kummer}}, \bibinfo {author} {\bibfnamefont {C.}~\bibnamefont {Nistor}},
  \bibinfo {author} {\bibfnamefont {P.}~\bibnamefont {Gambardella}},\ and\
  \bibinfo {author} {\bibfnamefont {H.}~\bibnamefont {Brune}},\ }\href
  {https://doi.org/10.1126/science.aad9898} {\bibfield  {journal} {\bibinfo
  {journal} {Science}\ }\textbf {\bibinfo {volume} {352}},\ \bibinfo {pages}
  {318} (\bibinfo {year} {2016})}\BibitemShut {NoStop}%
\bibitem [{\citenamefont {Baltic}\ \emph {et~al.}(2016)\citenamefont {Baltic},
  \citenamefont {Pivetta}, \citenamefont {Donati}, \citenamefont
  {W\"{a}ckerlin}, \citenamefont {Singha}, \citenamefont {Dreiser},
  \citenamefont {Rusponi},\ and\ \citenamefont
  {Brune}}]{baltic2016superlattice}%
  \BibitemOpen
  \bibfield  {author} {\bibinfo {author} {\bibfnamefont {R.}~\bibnamefont
  {Baltic}}, \bibinfo {author} {\bibfnamefont {M.}~\bibnamefont {Pivetta}},
  \bibinfo {author} {\bibfnamefont {F.}~\bibnamefont {Donati}}, \bibinfo
  {author} {\bibfnamefont {C.}~\bibnamefont {W\"{a}ckerlin}}, \bibinfo {author}
  {\bibfnamefont {A.}~\bibnamefont {Singha}}, \bibinfo {author} {\bibfnamefont
  {J.}~\bibnamefont {Dreiser}}, \bibinfo {author} {\bibfnamefont
  {S.}~\bibnamefont {Rusponi}},\ and\ \bibinfo {author} {\bibfnamefont
  {H.}~\bibnamefont {Brune}},\ }\href
  {https://doi.org/10.1021/acs.nanolett.6b03543} {\bibfield  {journal}
  {\bibinfo  {journal} {Nano letters}\ }\textbf {\bibinfo {volume} {16}},\
  \bibinfo {pages} {7610} (\bibinfo {year} {2016})}\BibitemShut {NoStop}%
\bibitem [{\citenamefont {Miyamachi}\ \emph {et~al.}(2013)\citenamefont
  {Miyamachi}, \citenamefont {Schuh}, \citenamefont {M{\"a}rkl}, \citenamefont
  {Bresch}, \citenamefont {St{\"o}hr}, \citenamefont {Karlewski}, \citenamefont
  {Andr{\'e}}, \citenamefont {Marthaler}, \citenamefont {Hoffmann},
  \citenamefont {Geilhufe}, \citenamefont {Hergert}, \citenamefont {Mertig},
  \citenamefont {Sch{\"o}n}, \citenamefont {Ernst},\ and\ \citenamefont
  {Wulfhekel}}]{miyamachi2013stabilizing}%
  \BibitemOpen
  \bibfield  {author} {\bibinfo {author} {\bibfnamefont {T.}~\bibnamefont
  {Miyamachi}}, \bibinfo {author} {\bibfnamefont {T.}~\bibnamefont {Schuh}},
  \bibinfo {author} {\bibfnamefont {T.}~\bibnamefont {M{\"a}rkl}}, \bibinfo
  {author} {\bibfnamefont {T.}~\bibnamefont {Bresch}, \bibfnamefont
  {C.and~Balashov}}, \bibinfo {author} {\bibfnamefont {A.}~\bibnamefont
  {St{\"o}hr}}, \bibinfo {author} {\bibfnamefont {C.}~\bibnamefont
  {Karlewski}}, \bibinfo {author} {\bibfnamefont {S.}~\bibnamefont
  {Andr{\'e}}}, \bibinfo {author} {\bibfnamefont {M.}~\bibnamefont
  {Marthaler}}, \bibinfo {author} {\bibfnamefont {M.}~\bibnamefont {Hoffmann}},
  \bibinfo {author} {\bibfnamefont {S.}~\bibnamefont {Geilhufe}, \bibfnamefont
  {M.~ans~Ostanin}}, \bibinfo {author} {\bibfnamefont {W.}~\bibnamefont
  {Hergert}}, \bibinfo {author} {\bibfnamefont {I.}~\bibnamefont {Mertig}},
  \bibinfo {author} {\bibfnamefont {G.}~\bibnamefont {Sch{\"o}n}}, \bibinfo
  {author} {\bibfnamefont {A.}~\bibnamefont {Ernst}},\ and\ \bibinfo {author}
  {\bibfnamefont {W.}~\bibnamefont {Wulfhekel}},\ }\href
  {https://doi.org/10.1038/nature12759} {\bibfield  {journal} {\bibinfo
  {journal} {Nature}\ }\textbf {\bibinfo {volume} {503}},\ \bibinfo {pages}
  {242} (\bibinfo {year} {2013})}\BibitemShut {NoStop}%
\bibitem [{\citenamefont {Ibanez-Azpiroz}\ \emph {et~al.}(2016)\citenamefont
  {Ibanez-Azpiroz}, \citenamefont {dos Santos~Dias}, \citenamefont {Blügel},\
  and\ \citenamefont {Lounis}}]{ibanez2016zero}%
  \BibitemOpen
  \bibfield  {author} {\bibinfo {author} {\bibfnamefont {J.}~\bibnamefont
  {Ibanez-Azpiroz}}, \bibinfo {author} {\bibfnamefont {M.}~\bibnamefont {dos
  Santos~Dias}}, \bibinfo {author} {\bibfnamefont {S.}~\bibnamefont
  {Blügel}},\ and\ \bibinfo {author} {\bibfnamefont {S.}~\bibnamefont
  {Lounis}},\ }\href {https://doi.org/10.1021/acs.nanolett.6b01344} {\bibfield
  {journal} {\bibinfo  {journal} {Nano letters}\ }\textbf {\bibinfo {volume}
  {16}},\ \bibinfo {pages} {4305} (\bibinfo {year} {2016})}\BibitemShut
  {NoStop}%
\bibitem [{\citenamefont {Bouaziz}\ \emph {et~al.}(2020)\citenamefont
  {Bouaziz}, \citenamefont {Ibanez-Azpiroz}, \citenamefont {Guimar\~aes},\ and\
  \citenamefont {Lounis}}]{Bouaziz2020}%
  \BibitemOpen
  \bibfield  {author} {\bibinfo {author} {\bibfnamefont {J.}~\bibnamefont
  {Bouaziz}}, \bibinfo {author} {\bibfnamefont {J.}~\bibnamefont
  {Ibanez-Azpiroz}}, \bibinfo {author} {\bibfnamefont {F.~S.~M.}\ \bibnamefont
  {Guimar\~aes}},\ and\ \bibinfo {author} {\bibfnamefont {S.}~\bibnamefont
  {Lounis}},\ }\href {https://doi.org/10.1103/PhysRevResearch.2.043357}
  {\bibfield  {journal} {\bibinfo  {journal} {Phys. Rev. Research}\ }\textbf
  {\bibinfo {volume} {2}},\ \bibinfo {pages} {043357} (\bibinfo {year}
  {2020})}\BibitemShut {NoStop}%
\bibitem [{\citenamefont {Grimm}\ \emph {et~al.}(2021)\citenamefont {Grimm},
  \citenamefont {Beckert}, \citenamefont {Aeppli},\ and\ \citenamefont
  {M\"{u}ller}}]{PRXQuantum.2.010312}%
  \BibitemOpen
  \bibfield  {author} {\bibinfo {author} {\bibfnamefont {M.}~\bibnamefont
  {Grimm}}, \bibinfo {author} {\bibfnamefont {A.}~\bibnamefont {Beckert}},
  \bibinfo {author} {\bibfnamefont {G.}~\bibnamefont {Aeppli}},\ and\ \bibinfo
  {author} {\bibfnamefont {M.}~\bibnamefont {M\"{u}ller}},\ }\href
  {https://doi.org/10.1103/PRXQuantum.2.010312} {\bibfield  {journal} {\bibinfo
   {journal} {PRX Quantum}\ }\textbf {\bibinfo {volume} {2}},\ \bibinfo {pages}
  {010312} (\bibinfo {year} {2021})}\BibitemShut {NoStop}%
\bibitem [{\citenamefont {Bertaina}\ \emph {et~al.}(2007)\citenamefont
  {Bertaina}, \citenamefont {Gambarelli}, \citenamefont {Tkachuk},
  \citenamefont {Kurkin}, \citenamefont {Malkin}, \citenamefont {Stepanov},\
  and\ \citenamefont {Barbara}}]{bertaina2007rare}%
  \BibitemOpen
  \bibfield  {author} {\bibinfo {author} {\bibfnamefont {S.}~\bibnamefont
  {Bertaina}}, \bibinfo {author} {\bibfnamefont {S.}~\bibnamefont
  {Gambarelli}}, \bibinfo {author} {\bibfnamefont {A.}~\bibnamefont {Tkachuk}},
  \bibinfo {author} {\bibfnamefont {I.}~\bibnamefont {Kurkin}}, \bibinfo
  {author} {\bibfnamefont {B.}~\bibnamefont {Malkin}}, \bibinfo {author}
  {\bibfnamefont {A.}~\bibnamefont {Stepanov}},\ and\ \bibinfo {author}
  {\bibfnamefont {B.}~\bibnamefont {Barbara}},\ }\href
  {https://doi.org/10.1038/nnano.2006.174} {\bibfield  {journal} {\bibinfo
  {journal} {Nature nanotechnology}\ }\textbf {\bibinfo {volume} {2}},\
  \bibinfo {pages} {39} (\bibinfo {year} {2007})}\BibitemShut {NoStop}%
\bibitem [{\citenamefont {Huttmann}\ \emph {et~al.}(2017)\citenamefont
  {Huttmann}, \citenamefont {Klar}, \citenamefont {Atodiresei}, \citenamefont
  {Schmitz-Antoniak}, \citenamefont {Smekhova}, \citenamefont
  {Martinez-Galera}, \citenamefont {Caciuc}, \citenamefont {Bihlmayer},
  \citenamefont {Bl{\"u}gel}, \citenamefont {Michely} \emph
  {et~al.}}]{3d4fHuttman}%
  \BibitemOpen
  \bibfield  {author} {\bibinfo {author} {\bibfnamefont {F.}~\bibnamefont
  {Huttmann}}, \bibinfo {author} {\bibfnamefont {D.}~\bibnamefont {Klar}},
  \bibinfo {author} {\bibfnamefont {N.}~\bibnamefont {Atodiresei}}, \bibinfo
  {author} {\bibfnamefont {C.}~\bibnamefont {Schmitz-Antoniak}}, \bibinfo
  {author} {\bibfnamefont {A.}~\bibnamefont {Smekhova}}, \bibinfo {author}
  {\bibfnamefont {A.~J.}\ \bibnamefont {Martinez-Galera}}, \bibinfo {author}
  {\bibfnamefont {V.}~\bibnamefont {Caciuc}}, \bibinfo {author} {\bibfnamefont
  {G.}~\bibnamefont {Bihlmayer}}, \bibinfo {author} {\bibfnamefont
  {S.}~\bibnamefont {Bl{\"u}gel}}, \bibinfo {author} {\bibfnamefont
  {T.}~\bibnamefont {Michely}}, \emph {et~al.},\ }\href
  {https://doi.org/10.1103/PhysRevB.95.075427} {\bibfield  {journal} {\bibinfo
  {journal} {Phys. Rev. B}\ }\textbf {\bibinfo {volume} {95}},\ \bibinfo
  {pages} {075427} (\bibinfo {year} {2017})}\BibitemShut {NoStop}%
\bibitem [{\citenamefont {Pivetta}\ \emph {et~al.}(2020)\citenamefont
  {Pivetta}, \citenamefont {Patthey}, \citenamefont {Di~Marco}, \citenamefont
  {Subramonian}, \citenamefont {Eriksson}, \citenamefont {Rusponi},\ and\
  \citenamefont {Brune}}]{Pivettameasure}%
  \BibitemOpen
  \bibfield  {author} {\bibinfo {author} {\bibfnamefont {M.}~\bibnamefont
  {Pivetta}}, \bibinfo {author} {\bibfnamefont {F.}~\bibnamefont {Patthey}},
  \bibinfo {author} {\bibfnamefont {I.}~\bibnamefont {Di~Marco}}, \bibinfo
  {author} {\bibfnamefont {A.}~\bibnamefont {Subramonian}}, \bibinfo {author}
  {\bibfnamefont {O.}~\bibnamefont {Eriksson}}, \bibinfo {author}
  {\bibfnamefont {S.}~\bibnamefont {Rusponi}},\ and\ \bibinfo {author}
  {\bibfnamefont {H.}~\bibnamefont {Brune}},\ }\href
  {https://doi.org/10.1103/PhysRevX.10.031054} {\bibfield  {journal} {\bibinfo
  {journal} {Phys. Rev. X}\ }\textbf {\bibinfo {volume} {10}},\ \bibinfo
  {pages} {031054} (\bibinfo {year} {2020})}\BibitemShut {NoStop}%
\bibitem [{\citenamefont {Liu}\ \emph {et~al.}(2010)\citenamefont {Liu},
  \citenamefont {Wang}, \citenamefont {Hupalo}, \citenamefont {Yao},
  \citenamefont {Tringides}, \citenamefont {Lu},\ and\ \citenamefont
  {Ho}}]{Liu2010}%
  \BibitemOpen
  \bibfield  {author} {\bibinfo {author} {\bibfnamefont {X.}~\bibnamefont
  {Liu}}, \bibinfo {author} {\bibfnamefont {C.~Z.}\ \bibnamefont {Wang}},
  \bibinfo {author} {\bibfnamefont {M.}~\bibnamefont {Hupalo}}, \bibinfo
  {author} {\bibfnamefont {Y.~X.}\ \bibnamefont {Yao}}, \bibinfo {author}
  {\bibfnamefont {M.~C.}\ \bibnamefont {Tringides}}, \bibinfo {author}
  {\bibfnamefont {W.~C.}\ \bibnamefont {Lu}},\ and\ \bibinfo {author}
  {\bibfnamefont {K.~M.}\ \bibnamefont {Ho}},\ }\href
  {https://doi.org/10.1103/PhysRevB.82.245408} {\bibfield  {journal} {\bibinfo
  {journal} {Physical Review B}\ }\textbf {\bibinfo {volume} {82}},\ \bibinfo
  {pages} {245408} (\bibinfo {year} {2010})}\BibitemShut {NoStop}%
\bibitem [{\citenamefont {Nistor}\ \emph {et~al.}(2014)\citenamefont {Nistor},
  \citenamefont {Mugarza}, \citenamefont {Stepanow}, \citenamefont
  {Gambardella}, \citenamefont {Kummer}, \citenamefont {Diez-Ferrer},
  \citenamefont {Coffey}, \citenamefont {{de la Fuente}}, \citenamefont
  {Ciria},\ and\ \citenamefont {Arnaudas}}]{Tm/W}%
  \BibitemOpen
  \bibfield  {author} {\bibinfo {author} {\bibfnamefont {C.}~\bibnamefont
  {Nistor}}, \bibinfo {author} {\bibfnamefont {A.}~\bibnamefont {Mugarza}},
  \bibinfo {author} {\bibfnamefont {S.}~\bibnamefont {Stepanow}}, \bibinfo
  {author} {\bibfnamefont {P.}~\bibnamefont {Gambardella}}, \bibinfo {author}
  {\bibfnamefont {K.}~\bibnamefont {Kummer}}, \bibinfo {author} {\bibfnamefont
  {J.~L.}\ \bibnamefont {Diez-Ferrer}}, \bibinfo {author} {\bibfnamefont
  {D.}~\bibnamefont {Coffey}}, \bibinfo {author} {\bibfnamefont
  {C.}~\bibnamefont {{de la Fuente}}}, \bibinfo {author} {\bibfnamefont
  {M.}~\bibnamefont {Ciria}},\ and\ \bibinfo {author} {\bibfnamefont {J.~I.}\
  \bibnamefont {Arnaudas}},\ }\href
  {https://doi.org/10.1103/PhysRevB.90.064423} {\bibfield  {journal} {\bibinfo
  {journal} {Phys. Rev. B}\ }\textbf {\bibinfo {volume} {90}},\ \bibinfo
  {pages} {064423} (\bibinfo {year} {2014})}\BibitemShut {NoStop}%
\bibitem [{\citenamefont {Bellini}\ \emph {et~al.}(2022)\citenamefont
  {Bellini}, \citenamefont {Rusponi}, \citenamefont {Koloren\u{c}},
  \citenamefont {Mahatha}, \citenamefont {Valbuena}, \citenamefont
  {Persichetti}, \citenamefont {Pivetta}, \citenamefont {Sorokin},
  \citenamefont {Merk}, \citenamefont {Reynaud}, \citenamefont {Sblendorio},
  \citenamefont {Stepanow}, \citenamefont {Nistor}, \citenamefont {Gargiani},
  \citenamefont {Betto}, \citenamefont {Mugarza}, \citenamefont {Gambardella},
  \citenamefont {Brune}, \citenamefont {Carbone},\ and\ \citenamefont
  {Barla}}]{DySTO}%
  \BibitemOpen
  \bibfield  {author} {\bibinfo {author} {\bibfnamefont {V.}~\bibnamefont
  {Bellini}}, \bibinfo {author} {\bibfnamefont {S.}~\bibnamefont {Rusponi}},
  \bibinfo {author} {\bibfnamefont {J.}~\bibnamefont {Koloren\u{c}}}, \bibinfo
  {author} {\bibfnamefont {S.~K.}\ \bibnamefont {Mahatha}}, \bibinfo {author}
  {\bibfnamefont {M.~A.}\ \bibnamefont {Valbuena}}, \bibinfo {author}
  {\bibfnamefont {L.}~\bibnamefont {Persichetti}}, \bibinfo {author}
  {\bibfnamefont {M.}~\bibnamefont {Pivetta}}, \bibinfo {author} {\bibfnamefont
  {B.~V.}\ \bibnamefont {Sorokin}}, \bibinfo {author} {\bibfnamefont
  {D.}~\bibnamefont {Merk}}, \bibinfo {author} {\bibfnamefont {S.}~\bibnamefont
  {Reynaud}}, \bibinfo {author} {\bibfnamefont {D.}~\bibnamefont {Sblendorio}},
  \bibinfo {author} {\bibfnamefont {S.}~\bibnamefont {Stepanow}}, \bibinfo
  {author} {\bibfnamefont {C.}~\bibnamefont {Nistor}}, \bibinfo {author}
  {\bibfnamefont {P.}~\bibnamefont {Gargiani}}, \bibinfo {author}
  {\bibfnamefont {D.}~\bibnamefont {Betto}}, \bibinfo {author} {\bibfnamefont
  {A.}~\bibnamefont {Mugarza}}, \bibinfo {author} {\bibfnamefont
  {P.}~\bibnamefont {Gambardella}}, \bibinfo {author} {\bibfnamefont
  {H.}~\bibnamefont {Brune}}, \bibinfo {author} {\bibfnamefont
  {C.}~\bibnamefont {Carbone}},\ and\ \bibinfo {author} {\bibfnamefont
  {A.}~\bibnamefont {Barla}},\ }\href {https://doi.org/10.1021/acsnano.2c04048}
  {\bibfield  {journal} {\bibinfo  {journal} {ACS Nano}\ }\textbf {\bibinfo
  {volume} {16}},\ \bibinfo {pages} {11182} (\bibinfo {year}
  {2022})}\BibitemShut {NoStop}%
\bibitem [{\citenamefont {Singha}\ \emph {et~al.}(2017)\citenamefont {Singha},
  \citenamefont {Baltic}, \citenamefont {Donati}, \citenamefont {W\"ackerlin},
  \citenamefont {Dreiser}, \citenamefont {Persichetti}, \citenamefont
  {Stepanow}, \citenamefont {Gambardella}, \citenamefont {Rusponi},\ and\
  \citenamefont {Brune}}]{4focc}%
  \BibitemOpen
  \bibfield  {author} {\bibinfo {author} {\bibfnamefont {A.}~\bibnamefont
  {Singha}}, \bibinfo {author} {\bibfnamefont {R.}~\bibnamefont {Baltic}},
  \bibinfo {author} {\bibfnamefont {F.}~\bibnamefont {Donati}}, \bibinfo
  {author} {\bibfnamefont {C.}~\bibnamefont {W\"ackerlin}}, \bibinfo {author}
  {\bibfnamefont {J.}~\bibnamefont {Dreiser}}, \bibinfo {author} {\bibfnamefont
  {L.}~\bibnamefont {Persichetti}}, \bibinfo {author} {\bibfnamefont
  {S.}~\bibnamefont {Stepanow}}, \bibinfo {author} {\bibfnamefont
  {P.}~\bibnamefont {Gambardella}}, \bibinfo {author} {\bibfnamefont
  {S.}~\bibnamefont {Rusponi}},\ and\ \bibinfo {author} {\bibfnamefont
  {H.}~\bibnamefont {Brune}},\ }\href
  {https://doi.org/10.1103/PhysRevB.96.224418} {\bibfield  {journal} {\bibinfo
  {journal} {Phys. Rev. B}\ }\textbf {\bibinfo {volume} {96}},\ \bibinfo
  {pages} {224418} (\bibinfo {year} {2017})}\BibitemShut {NoStop}%
\bibitem [{\citenamefont {Baltic}\ \emph {et~al.}(2018)\citenamefont {Baltic},
  \citenamefont {Donati}, \citenamefont {Singha}, \citenamefont
  {W{\"a}ckerlin}, \citenamefont {Dreiser}, \citenamefont {Delley},
  \citenamefont {Pivetta}, \citenamefont {Rusponi},\ and\ \citenamefont
  {Brune}}]{Baltic2018}%
  \BibitemOpen
  \bibfield  {author} {\bibinfo {author} {\bibfnamefont {R.}~\bibnamefont
  {Baltic}}, \bibinfo {author} {\bibfnamefont {F.}~\bibnamefont {Donati}},
  \bibinfo {author} {\bibfnamefont {A.}~\bibnamefont {Singha}}, \bibinfo
  {author} {\bibfnamefont {C.}~\bibnamefont {W{\"a}ckerlin}}, \bibinfo {author}
  {\bibfnamefont {J.}~\bibnamefont {Dreiser}}, \bibinfo {author} {\bibfnamefont
  {B.}~\bibnamefont {Delley}}, \bibinfo {author} {\bibfnamefont
  {M.}~\bibnamefont {Pivetta}}, \bibinfo {author} {\bibfnamefont
  {S.}~\bibnamefont {Rusponi}},\ and\ \bibinfo {author} {\bibfnamefont
  {H.}~\bibnamefont {Brune}},\ }\href
  {https://doi.org/10.1103/PhysRevB.98.024412} {\bibfield  {journal} {\bibinfo
  {journal} {Phys. Rev. B}\ }\textbf {\bibinfo {volume} {98}},\ \bibinfo
  {pages} {024412} (\bibinfo {year} {2018})}\BibitemShut {NoStop}%
\bibitem [{\citenamefont {Donati}\ \emph {et~al.}(2014)\citenamefont {Donati},
  \citenamefont {Singha}, \citenamefont {Stepanow}, \citenamefont
  {W\"ackerlin}, \citenamefont {Dreiser}, \citenamefont {Gambardella},
  \citenamefont {Rusponi},\ and\ \citenamefont {Brune}}]{BruneHoEr}%
  \BibitemOpen
  \bibfield  {author} {\bibinfo {author} {\bibfnamefont {F.}~\bibnamefont
  {Donati}}, \bibinfo {author} {\bibfnamefont {A.}~\bibnamefont {Singha}},
  \bibinfo {author} {\bibfnamefont {S.}~\bibnamefont {Stepanow}}, \bibinfo
  {author} {\bibfnamefont {C.}~\bibnamefont {W\"ackerlin}}, \bibinfo {author}
  {\bibfnamefont {J.}~\bibnamefont {Dreiser}}, \bibinfo {author} {\bibfnamefont
  {P.}~\bibnamefont {Gambardella}}, \bibinfo {author} {\bibfnamefont
  {S.}~\bibnamefont {Rusponi}},\ and\ \bibinfo {author} {\bibfnamefont
  {H.}~\bibnamefont {Brune}},\ }\href
  {https://doi.org/10.1103/PhysRevLett.113.237201} {\bibfield  {journal}
  {\bibinfo  {journal} {Phys. Rev. Lett.}\ }\textbf {\bibinfo {volume} {113}},\
  \bibinfo {pages} {237201} (\bibinfo {year} {2014})}\BibitemShut {NoStop}%
\bibitem [{\citenamefont {Donati}\ \emph {et~al.}(2021)\citenamefont {Donati},
  \citenamefont {Pivetta}, \citenamefont {Wolf}, \citenamefont {Singha},
  \citenamefont {Wäckerlin}, \citenamefont {Baltic}, \citenamefont
  {Fernandes}, \citenamefont {de~Groot}, \citenamefont {Ahmed}, \citenamefont
  {Persichetti}, \citenamefont {Nistor}, \citenamefont {Dreiser}, \citenamefont
  {Barla}, \citenamefont {Gambardella}, \citenamefont {Brune},\ and\
  \citenamefont {Rusponi}}]{Rusponithickness}%
  \BibitemOpen
  \bibfield  {author} {\bibinfo {author} {\bibfnamefont {F.}~\bibnamefont
  {Donati}}, \bibinfo {author} {\bibfnamefont {M.}~\bibnamefont {Pivetta}},
  \bibinfo {author} {\bibfnamefont {C.}~\bibnamefont {Wolf}}, \bibinfo {author}
  {\bibfnamefont {A.}~\bibnamefont {Singha}}, \bibinfo {author} {\bibfnamefont
  {C.}~\bibnamefont {Wäckerlin}}, \bibinfo {author} {\bibfnamefont
  {R.}~\bibnamefont {Baltic}}, \bibinfo {author} {\bibfnamefont
  {E.}~\bibnamefont {Fernandes}}, \bibinfo {author} {\bibfnamefont {J.-G.}\
  \bibnamefont {de~Groot}}, \bibinfo {author} {\bibfnamefont {S.~L.}\
  \bibnamefont {Ahmed}}, \bibinfo {author} {\bibfnamefont {L.}~\bibnamefont
  {Persichetti}}, \bibinfo {author} {\bibfnamefont {C.}~\bibnamefont {Nistor}},
  \bibinfo {author} {\bibfnamefont {J.}~\bibnamefont {Dreiser}}, \bibinfo
  {author} {\bibfnamefont {A.}~\bibnamefont {Barla}}, \bibinfo {author}
  {\bibfnamefont {P.}~\bibnamefont {Gambardella}}, \bibinfo {author}
  {\bibfnamefont {H.}~\bibnamefont {Brune}},\ and\ \bibinfo {author}
  {\bibfnamefont {S.}~\bibnamefont {Rusponi}},\ }\href
  {https://doi.org/10.1021/acs.nanolett.1c02744} {\bibfield  {journal}
  {\bibinfo  {journal} {Nano Letters}\ }\textbf {\bibinfo {volume} {21}},\
  \bibinfo {pages} {8266} (\bibinfo {year} {2021})}\BibitemShut {NoStop}%
\bibitem [{\citenamefont {Herman}\ \emph {et~al.}(2022)\citenamefont {Herman},
  \citenamefont {Kraus}, \citenamefont {Tsukamoto}, \citenamefont {Spieker},
  \citenamefont {Caciuc}, \citenamefont {Lojewski}, \citenamefont
  {G{\"u}nzing}, \citenamefont {Dreiser}, \citenamefont {Delley}, \citenamefont
  {Ollefs}, \citenamefont {Michely}, \citenamefont {Atodiresei},\ and\
  \citenamefont {Wende}}]{herman2022tailoring}%
  \BibitemOpen
  \bibfield  {author} {\bibinfo {author} {\bibfnamefont {A.}~\bibnamefont
  {Herman}}, \bibinfo {author} {\bibfnamefont {S.}~\bibnamefont {Kraus}},
  \bibinfo {author} {\bibfnamefont {S.}~\bibnamefont {Tsukamoto}}, \bibinfo
  {author} {\bibfnamefont {L.}~\bibnamefont {Spieker}}, \bibinfo {author}
  {\bibfnamefont {V.}~\bibnamefont {Caciuc}}, \bibinfo {author} {\bibfnamefont
  {T.}~\bibnamefont {Lojewski}}, \bibinfo {author} {\bibfnamefont
  {D.}~\bibnamefont {G{\"u}nzing}}, \bibinfo {author} {\bibfnamefont
  {J.}~\bibnamefont {Dreiser}}, \bibinfo {author} {\bibfnamefont
  {B.}~\bibnamefont {Delley}}, \bibinfo {author} {\bibfnamefont
  {K.}~\bibnamefont {Ollefs}}, \bibinfo {author} {\bibfnamefont
  {T.}~\bibnamefont {Michely}}, \bibinfo {author} {\bibfnamefont
  {N.}~\bibnamefont {Atodiresei}},\ and\ \bibinfo {author} {\bibfnamefont
  {H.}~\bibnamefont {Wende}},\ }\href {https://doi.org/10.1039/d2nr01458k}
  {\bibfield  {journal} {\bibinfo  {journal} {Nanoscale}\ }\textbf {\bibinfo
  {volume} {14}},\ \bibinfo {pages} {7682} (\bibinfo {year}
  {2022})}\BibitemShut {NoStop}%
\bibitem [{\citenamefont {Peters}\ \emph {et~al.}(2014)\citenamefont {Peters},
  \citenamefont {Di~Marco}, \citenamefont {Thunstr\"om}, \citenamefont
  {Katsnelson}, \citenamefont {Kirilyuk},\ and\ \citenamefont
  {Eriksson}}]{Peters2014}%
  \BibitemOpen
  \bibfield  {author} {\bibinfo {author} {\bibfnamefont {L.}~\bibnamefont
  {Peters}}, \bibinfo {author} {\bibfnamefont {I.}~\bibnamefont {Di~Marco}},
  \bibinfo {author} {\bibfnamefont {P.}~\bibnamefont {Thunstr\"om}}, \bibinfo
  {author} {\bibfnamefont {M.~I.}\ \bibnamefont {Katsnelson}}, \bibinfo
  {author} {\bibfnamefont {A.}~\bibnamefont {Kirilyuk}},\ and\ \bibinfo
  {author} {\bibfnamefont {O.}~\bibnamefont {Eriksson}},\ }\href
  {https://doi.org/10.1103/PhysRevB.89.205109} {\bibfield  {journal} {\bibinfo
  {journal} {Phys. Rev. B}\ }\textbf {\bibinfo {volume} {89}},\ \bibinfo
  {pages} {205109} (\bibinfo {year} {2014})}\BibitemShut {NoStop}%
\bibitem [{\citenamefont {Shick}\ \emph {et~al.}(2017)\citenamefont {Shick},
  \citenamefont {Lichtenstein}, \citenamefont {Shapiro},\ and\ \citenamefont
  {Kolorenc}}]{ShickHoPt(111)}%
  \BibitemOpen
  \bibfield  {author} {\bibinfo {author} {\bibfnamefont {A.~B.}\ \bibnamefont
  {Shick}}, \bibinfo {author} {\bibfnamefont {A.~I.}\ \bibnamefont
  {Lichtenstein}}, \bibinfo {author} {\bibfnamefont {D.~S.}\ \bibnamefont
  {Shapiro}},\ and\ \bibinfo {author} {\bibfnamefont {J.}~\bibnamefont
  {Kolorenc}},\ }\href {https://doi.org/10.1038/s41598-017-02809-7} {\bibfield
  {journal} {\bibinfo  {journal} {Scientific Reports}\ }\textbf {\bibinfo
  {volume} {7}},\ \bibinfo {pages} {3} (\bibinfo {year} {2017})}\BibitemShut
  {NoStop}%
\bibitem [{\citenamefont {Shick}\ and\ \citenamefont
  {Denisov}(2019)}]{SHICK2019}%
  \BibitemOpen
  \bibfield  {author} {\bibinfo {author} {\bibfnamefont {A.~B.}\ \bibnamefont
  {Shick}}\ and\ \bibinfo {author} {\bibfnamefont {A.~Y.}\ \bibnamefont
  {Denisov}},\ }\href
  {https://doi.org/https://doi.org/10.1016/j.jmmm.2018.11.078} {\bibfield
  {journal} {\bibinfo  {journal} {Journal of Magnetism and Magnetic Materials}\
  }\textbf {\bibinfo {volume} {475}},\ \bibinfo {pages} {211} (\bibinfo {year}
  {2019})}\BibitemShut {NoStop}%
\bibitem [{\citenamefont {Shick}\ \emph {et~al.}(2009)\citenamefont {Shick},
  \citenamefont {Koloren\v{c}}, \citenamefont {Lichtenstein},\ and\
  \citenamefont {Havela}}]{shick2009HIA}%
  \BibitemOpen
  \bibfield  {author} {\bibinfo {author} {\bibfnamefont {A.~B.}\ \bibnamefont
  {Shick}}, \bibinfo {author} {\bibfnamefont {J.}~\bibnamefont {Koloren\v{c}}},
  \bibinfo {author} {\bibfnamefont {A.~I.}\ \bibnamefont {Lichtenstein}},\ and\
  \bibinfo {author} {\bibfnamefont {L.}~\bibnamefont {Havela}},\ }\href
  {https://doi.org/10.1103/PhysRevB.80.085106} {\bibfield  {journal} {\bibinfo
  {journal} {Physical Review B}\ }\textbf {\bibinfo {volume} {80}},\ \bibinfo
  {pages} {085106} (\bibinfo {year} {2009})}\BibitemShut {NoStop}%
\bibitem [{\citenamefont {Held}(2007)}]{held2007electronic}%
  \BibitemOpen
  \bibfield  {author} {\bibinfo {author} {\bibfnamefont {K.}~\bibnamefont
  {Held}},\ }\href {https://doi.org/10.1080/00018730701619647} {\bibfield
  {journal} {\bibinfo  {journal} {Advances in Physics}\ }\textbf {\bibinfo
  {volume} {56}},\ \bibinfo {pages} {829} (\bibinfo {year} {2007})}\BibitemShut
  {NoStop}%
\bibitem [{\citenamefont {Anisimov}\ \emph {et~al.}(1997)\citenamefont
  {Anisimov}, \citenamefont {Aryasetiawan},\ and\ \citenamefont
  {Lichtenstein}}]{Anisimov:1997}%
  \BibitemOpen
  \bibfield  {author} {\bibinfo {author} {\bibfnamefont {V.~I.}\ \bibnamefont
  {Anisimov}}, \bibinfo {author} {\bibfnamefont {F.}~\bibnamefont
  {Aryasetiawan}},\ and\ \bibinfo {author} {\bibfnamefont {A.}~\bibnamefont
  {Lichtenstein}},\ }\href {https://doi.org/10.1088/0953-8984/9/4/002}
  {\bibfield  {journal} {\bibinfo  {journal} {J. Phys. Condens. Matter}\
  }\textbf {\bibinfo {volume} {9}},\ \bibinfo {pages} {767} (\bibinfo {year}
  {1997})}\BibitemShut {NoStop}%
\bibitem [{\citenamefont {Pivetta}\ \emph {et~al.}(2018)\citenamefont
  {Pivetta}, \citenamefont {Rusponi},\ and\ \citenamefont
  {Brune}}]{Pivetta2018}%
  \BibitemOpen
  \bibfield  {author} {\bibinfo {author} {\bibfnamefont {M.}~\bibnamefont
  {Pivetta}}, \bibinfo {author} {\bibfnamefont {S.}~\bibnamefont {Rusponi}},\
  and\ \bibinfo {author} {\bibfnamefont {H.}~\bibnamefont {Brune}},\ }\href
  {https://doi.org/10.1103/PhysRevB.98.115417} {\bibfield  {journal} {\bibinfo
  {journal} {Phys. Rev. B}\ }\textbf {\bibinfo {volume} {98}},\ \bibinfo
  {pages} {115417} (\bibinfo {year} {2018})}\BibitemShut {NoStop}%
\bibitem [{\citenamefont {Shick}\ \emph
  {et~al.}(1999{\natexlab{a}})\citenamefont {Shick}, \citenamefont
  {Liechtenstein},\ and\ \citenamefont {Pickett}}]{Shick1999}%
  \BibitemOpen
  \bibfield  {author} {\bibinfo {author} {\bibfnamefont {A.~B.}\ \bibnamefont
  {Shick}}, \bibinfo {author} {\bibfnamefont {A.~I.}\ \bibnamefont
  {Liechtenstein}},\ and\ \bibinfo {author} {\bibfnamefont {W.~E.}\
  \bibnamefont {Pickett}},\ }\href {https://doi.org/10.1103/PhysRevB.60.10763}
  {\bibfield  {journal} {\bibinfo  {journal} {Phys. Rev. B}\ }\textbf {\bibinfo
  {volume} {60}},\ \bibinfo {pages} {10763} (\bibinfo {year}
  {1999}{\natexlab{a}})}\BibitemShut {NoStop}%
\bibitem [{FLE()}]{FLEUR}%
  \BibitemOpen
  \href@noop {} {}\bibinfo {howpublished}
  {\url{https://www.flapw.de/}}\BibitemShut {NoStop}%
\bibitem [{\citenamefont {Wortmann}\ \emph {et~al.}(2023)\citenamefont
  {Wortmann}, \citenamefont {Michalicek}, \citenamefont {Baadji}, \citenamefont
  {Betzinger}, \citenamefont {Bihlmayer}, \citenamefont {Br{\"o}der},
  \citenamefont {Burnus}, \citenamefont {Enkovaara}, \citenamefont {Freimuth},
  \citenamefont {Friedrich}, \citenamefont {Gerhorst}, \citenamefont
  {Granberg~Cauchi}, \citenamefont {Grytsiuk}, \citenamefont {Hanke},
  \citenamefont {Hanke}, \citenamefont {Heide}, \citenamefont {Heinze},
  \citenamefont {Hilgers}, \citenamefont {Janssen}, \citenamefont
  {Kl{\"u}ppelberg}, \citenamefont {Kovacik}, \citenamefont {Kurz},
  \citenamefont {Lezaic}, \citenamefont {Madsen}, \citenamefont {Mokrousov},
  \citenamefont {Neukirchen}, \citenamefont {Redies}, \citenamefont {Rost},
  \citenamefont {Schlipf}, \citenamefont {Schindlmayr}, \citenamefont
  {Winkelmann},\ and\ \citenamefont {Bl{\"u}gel}}]{wortmann2023fleur}%
  \BibitemOpen
  \bibfield  {author} {\bibinfo {author} {\bibfnamefont {D.}~\bibnamefont
  {Wortmann}}, \bibinfo {author} {\bibfnamefont {G.}~\bibnamefont
  {Michalicek}}, \bibinfo {author} {\bibfnamefont {N.}~\bibnamefont {Baadji}},
  \bibinfo {author} {\bibfnamefont {M.}~\bibnamefont {Betzinger}}, \bibinfo
  {author} {\bibfnamefont {G.}~\bibnamefont {Bihlmayer}}, \bibinfo {author}
  {\bibfnamefont {J.}~\bibnamefont {Br{\"o}der}}, \bibinfo {author}
  {\bibfnamefont {T.}~\bibnamefont {Burnus}}, \bibinfo {author} {\bibfnamefont
  {J.}~\bibnamefont {Enkovaara}}, \bibinfo {author} {\bibfnamefont
  {F.}~\bibnamefont {Freimuth}}, \bibinfo {author} {\bibfnamefont
  {C.}~\bibnamefont {Friedrich}}, \bibinfo {author} {\bibfnamefont {C.-R.}\
  \bibnamefont {Gerhorst}}, \bibinfo {author} {\bibfnamefont {S.}~\bibnamefont
  {Granberg~Cauchi}}, \bibinfo {author} {\bibfnamefont {U.}~\bibnamefont
  {Grytsiuk}}, \bibinfo {author} {\bibfnamefont {A.}~\bibnamefont {Hanke}},
  \bibinfo {author} {\bibfnamefont {J.-P.}\ \bibnamefont {Hanke}}, \bibinfo
  {author} {\bibfnamefont {M.}~\bibnamefont {Heide}}, \bibinfo {author}
  {\bibfnamefont {S.}~\bibnamefont {Heinze}}, \bibinfo {author} {\bibfnamefont
  {R.}~\bibnamefont {Hilgers}}, \bibinfo {author} {\bibfnamefont
  {H.}~\bibnamefont {Janssen}}, \bibinfo {author} {\bibfnamefont {D.~A.}\
  \bibnamefont {Kl{\"u}ppelberg}}, \bibinfo {author} {\bibfnamefont
  {R.}~\bibnamefont {Kovacik}}, \bibinfo {author} {\bibfnamefont
  {P.}~\bibnamefont {Kurz}}, \bibinfo {author} {\bibfnamefont {M.}~\bibnamefont
  {Lezaic}}, \bibinfo {author} {\bibfnamefont {G.~K.~H.}\ \bibnamefont
  {Madsen}}, \bibinfo {author} {\bibfnamefont {Y.}~\bibnamefont {Mokrousov}},
  \bibinfo {author} {\bibfnamefont {A.}~\bibnamefont {Neukirchen}}, \bibinfo
  {author} {\bibfnamefont {M.}~\bibnamefont {Redies}}, \bibinfo {author}
  {\bibfnamefont {S.}~\bibnamefont {Rost}}, \bibinfo {author} {\bibfnamefont
  {M.}~\bibnamefont {Schlipf}}, \bibinfo {author} {\bibfnamefont
  {A.}~\bibnamefont {Schindlmayr}}, \bibinfo {author} {\bibfnamefont
  {M.}~\bibnamefont {Winkelmann}},\ and\ \bibinfo {author} {\bibfnamefont
  {S.}~\bibnamefont {Bl{\"u}gel}},\ }\href
  {https://doi.org/10.5281/zenodo.7891361} {\emph {\bibinfo {title} {FLEUR}}},\
  \bibinfo {type} {Tech. Rep.}\ (\bibinfo {year} {2023})\BibitemShut {NoStop}%
\bibitem [{\citenamefont {Jugovac}\ \emph {et~al.}(2023)\citenamefont
  {Jugovac}, \citenamefont {Cojocariu}, \citenamefont {S{\'a}nchez-Barriga},
  \citenamefont {Gargiani}, \citenamefont {Valvidares}, \citenamefont {Feyer},
  \citenamefont {Bl{\"u}gel}, \citenamefont {Bihlmayer},\ and\ \citenamefont
  {Perna}}]{jugovac2023inducing}%
  \BibitemOpen
  \bibfield  {author} {\bibinfo {author} {\bibfnamefont {M.}~\bibnamefont
  {Jugovac}}, \bibinfo {author} {\bibfnamefont {I.}~\bibnamefont {Cojocariu}},
  \bibinfo {author} {\bibfnamefont {J.}~\bibnamefont {S{\'a}nchez-Barriga}},
  \bibinfo {author} {\bibfnamefont {P.}~\bibnamefont {Gargiani}}, \bibinfo
  {author} {\bibfnamefont {M.}~\bibnamefont {Valvidares}}, \bibinfo {author}
  {\bibfnamefont {V.}~\bibnamefont {Feyer}}, \bibinfo {author} {\bibfnamefont
  {S.}~\bibnamefont {Bl{\"u}gel}}, \bibinfo {author} {\bibfnamefont
  {G.}~\bibnamefont {Bihlmayer}},\ and\ \bibinfo {author} {\bibfnamefont
  {P.}~\bibnamefont {Perna}},\ }\href {https://doi.org/10.1002/adma.202301441}
  {\bibfield  {journal} {\bibinfo  {journal} {Advanced Materials}\ ,\ \bibinfo
  {pages} {2301441}} (\bibinfo {year} {2023})}\BibitemShut {NoStop}%
\bibitem [{\citenamefont {F{\"o}rster}\ \emph {et~al.}(2012)\citenamefont
  {F{\"o}rster}, \citenamefont {Wehling}, \citenamefont {Schumacher},
  \citenamefont {Rosch},\ and\ \citenamefont {Michely}}]{forster2012phase}%
  \BibitemOpen
  \bibfield  {author} {\bibinfo {author} {\bibfnamefont {D.~F.}\ \bibnamefont
  {F{\"o}rster}}, \bibinfo {author} {\bibfnamefont {T.~O.}\ \bibnamefont
  {Wehling}}, \bibinfo {author} {\bibfnamefont {S.}~\bibnamefont {Schumacher}},
  \bibinfo {author} {\bibfnamefont {A.}~\bibnamefont {Rosch}},\ and\ \bibinfo
  {author} {\bibfnamefont {T.}~\bibnamefont {Michely}},\ }\href
  {https://doi.org/10.1088/1367-2630/14/2/023022} {\bibfield  {journal}
  {\bibinfo  {journal} {New Journal of Physics}\ }\textbf {\bibinfo {volume}
  {14}},\ \bibinfo {pages} {023022} (\bibinfo {year} {2012})}\BibitemShut
  {NoStop}%
\bibitem [{\citenamefont {Li}\ \emph {et~al.}(1990)\citenamefont {Li},
  \citenamefont {Freeman}, \citenamefont {Jansen},\ and\ \citenamefont
  {Fu}}]{2nd_variation}%
  \BibitemOpen
  \bibfield  {author} {\bibinfo {author} {\bibfnamefont {C.}~\bibnamefont
  {Li}}, \bibinfo {author} {\bibfnamefont {A.~J.}\ \bibnamefont {Freeman}},
  \bibinfo {author} {\bibfnamefont {H.~J.~F.}\ \bibnamefont {Jansen}},\ and\
  \bibinfo {author} {\bibfnamefont {C.~L.}\ \bibnamefont {Fu}},\ }\href
  {https://doi.org/10.1103/PhysRevB.42.5433} {\bibfield  {journal} {\bibinfo
  {journal} {Phys. Rev. B}\ }\textbf {\bibinfo {volume} {42}},\ \bibinfo
  {pages} {5433} (\bibinfo {year} {1990})}\BibitemShut {NoStop}%
\bibitem [{\citenamefont {Perdew}\ \emph {et~al.}(1996)\citenamefont {Perdew},
  \citenamefont {Burke},\ and\ \citenamefont {Ernzerhof}}]{PBE}%
  \BibitemOpen
  \bibfield  {author} {\bibinfo {author} {\bibfnamefont {J.~P.}\ \bibnamefont
  {Perdew}}, \bibinfo {author} {\bibfnamefont {K.}~\bibnamefont {Burke}},\ and\
  \bibinfo {author} {\bibfnamefont {M.}~\bibnamefont {Ernzerhof}},\ }\href
  {https://doi.org/10.1103/PhysRevLett.77.3865} {\bibfield  {journal} {\bibinfo
   {journal} {Phys. Rev. Lett.}\ }\textbf {\bibinfo {volume} {77}},\ \bibinfo
  {pages} {3865} (\bibinfo {year} {1996})}\BibitemShut {NoStop}%
\bibitem [{\citenamefont {Shick}\ \emph
  {et~al.}(1999{\natexlab{b}})\citenamefont {Shick}, \citenamefont
  {Liechtenstein},\ and\ \citenamefont {Pickett}}]{DFTUShick}%
  \BibitemOpen
  \bibfield  {author} {\bibinfo {author} {\bibfnamefont {A.~B.}\ \bibnamefont
  {Shick}}, \bibinfo {author} {\bibfnamefont {A.~I.}\ \bibnamefont
  {Liechtenstein}},\ and\ \bibinfo {author} {\bibfnamefont {W.~E.}\
  \bibnamefont {Pickett}},\ }\href {https://doi.org/10.1103/PhysRevB.60.10763}
  {\bibfield  {journal} {\bibinfo  {journal} {Phys. Rev. B}\ }\textbf {\bibinfo
  {volume} {60}},\ \bibinfo {pages} {10763} (\bibinfo {year}
  {1999}{\natexlab{b}})}\BibitemShut {NoStop}%
\bibitem [{\citenamefont {van~der Marel}\ and\ \citenamefont
  {Sawatzky}(1988)}]{UandJ}%
  \BibitemOpen
  \bibfield  {author} {\bibinfo {author} {\bibfnamefont {D.}~\bibnamefont
  {van~der Marel}}\ and\ \bibinfo {author} {\bibfnamefont {G.~A.}\ \bibnamefont
  {Sawatzky}},\ }\href {https://doi.org/10.1103/PhysRevB.37.10674} {\bibfield
  {journal} {\bibinfo  {journal} {Phys. Rev. B}\ }\textbf {\bibinfo {volume}
  {37}},\ \bibinfo {pages} {10674} (\bibinfo {year} {1988})}\BibitemShut
  {NoStop}%
\bibitem [{\citenamefont {Locht}\ \emph {et~al.}(2016)\citenamefont {Locht},
  \citenamefont {Kvashnin}, \citenamefont {Rodrigues}, \citenamefont {Pereiro},
  \citenamefont {Bergman}, \citenamefont {Bergqvist}, \citenamefont
  {Lichtenstein}, \citenamefont {Katsnelson}, \citenamefont {Delin},
  \citenamefont {Klautau}, \citenamefont {Johansson}, \citenamefont
  {Di~Marco},\ and\ \citenamefont {Eriksson}}]{Locht2018}%
  \BibitemOpen
  \bibfield  {author} {\bibinfo {author} {\bibfnamefont {I.~L.~M.}\
  \bibnamefont {Locht}}, \bibinfo {author} {\bibfnamefont {Y.~O.}\ \bibnamefont
  {Kvashnin}}, \bibinfo {author} {\bibfnamefont {D.~C.~M.}\ \bibnamefont
  {Rodrigues}}, \bibinfo {author} {\bibfnamefont {M.}~\bibnamefont {Pereiro}},
  \bibinfo {author} {\bibfnamefont {A.}~\bibnamefont {Bergman}}, \bibinfo
  {author} {\bibfnamefont {L.}~\bibnamefont {Bergqvist}}, \bibinfo {author}
  {\bibfnamefont {A.~I.}\ \bibnamefont {Lichtenstein}}, \bibinfo {author}
  {\bibfnamefont {M.~I.}\ \bibnamefont {Katsnelson}}, \bibinfo {author}
  {\bibfnamefont {A.}~\bibnamefont {Delin}}, \bibinfo {author} {\bibfnamefont
  {A.~B.}\ \bibnamefont {Klautau}}, \bibinfo {author} {\bibfnamefont
  {B.}~\bibnamefont {Johansson}}, \bibinfo {author} {\bibfnamefont
  {I.}~\bibnamefont {Di~Marco}},\ and\ \bibinfo {author} {\bibfnamefont
  {O.}~\bibnamefont {Eriksson}},\ }\href
  {https://doi.org/10.1103/PhysRevB.94.085137} {\bibfield  {journal} {\bibinfo
  {journal} {Phys. Rev. B}\ }\textbf {\bibinfo {volume} {94}},\ \bibinfo
  {pages} {085137} (\bibinfo {year} {2016})}\BibitemShut {NoStop}%
\bibitem [{\citenamefont {Liu}\ \emph {et~al.}(2012)\citenamefont {Liu},
  \citenamefont {Wang}, \citenamefont {Hupalo}, \citenamefont {Lu},
  \citenamefont {Tringides}, \citenamefont {Yao},\ and\ \citenamefont
  {Ho}}]{Liu2012}%
  \BibitemOpen
  \bibfield  {author} {\bibinfo {author} {\bibfnamefont {X.}~\bibnamefont
  {Liu}}, \bibinfo {author} {\bibfnamefont {C.~Z.}\ \bibnamefont {Wang}},
  \bibinfo {author} {\bibfnamefont {M.}~\bibnamefont {Hupalo}}, \bibinfo
  {author} {\bibfnamefont {W.~C.}\ \bibnamefont {Lu}}, \bibinfo {author}
  {\bibfnamefont {M.~C.}\ \bibnamefont {Tringides}}, \bibinfo {author}
  {\bibfnamefont {Y.~X.}\ \bibnamefont {Yao}},\ and\ \bibinfo {author}
  {\bibfnamefont {K.~M.}\ \bibnamefont {Ho}},\ }\href
  {https://doi.org/10.1039/C2CP40527J} {\bibfield  {journal} {\bibinfo
  {journal} {Phys. Chem. Chem. Phys.}\ }\textbf {\bibinfo {volume} {14}},\
  \bibinfo {pages} {9157} (\bibinfo {year} {2012})}\BibitemShut {NoStop}%
\bibitem [{\citenamefont {Basiuk}\ \emph {et~al.}(2022)\citenamefont {Basiuk},
  \citenamefont {Prezhdo},\ and\ \citenamefont {Basiuk}}]{Basiuk2022}%
  \BibitemOpen
  \bibfield  {author} {\bibinfo {author} {\bibfnamefont {V.~A.}\ \bibnamefont
  {Basiuk}}, \bibinfo {author} {\bibfnamefont {O.~V.}\ \bibnamefont
  {Prezhdo}},\ and\ \bibinfo {author} {\bibfnamefont {E.~V.}\ \bibnamefont
  {Basiuk}},\ }\href {https://doi.org/10.1021/acs.jpclett.2c01580} {\bibfield
  {journal} {\bibinfo  {journal} {The Journal of Physical Chemistry Letters}\
  }\textbf {\bibinfo {volume} {13}},\ \bibinfo {pages} {6042} (\bibinfo {year}
  {2022})}\BibitemShut {NoStop}%
\bibitem [{\citenamefont {Hutchings}(1964)}]{hutchings1964point}%
  \BibitemOpen
  \bibfield  {author} {\bibinfo {author} {\bibfnamefont {M.~T.}\ \bibnamefont
  {Hutchings}},\ }in\ \href {https://doi.org/10.1016/S0081-1947(08)60517-2}
  {\emph {\bibinfo {booktitle} {Solid state physics}}},\ Vol.~\bibinfo {volume}
  {16}\ (\bibinfo  {publisher} {Elsevier},\ \bibinfo {year} {1964})\ pp.\
  \bibinfo {pages} {227--273}\BibitemShut {NoStop}%
\bibitem [{\citenamefont {Skomski}\ \emph {et~al.}(2020)\citenamefont
  {Skomski}, \citenamefont {Manchanda},\ and\ \citenamefont
  {Kashyap}}]{skomski2020anisotropy}%
  \BibitemOpen
  \bibfield  {author} {\bibinfo {author} {\bibfnamefont {R.}~\bibnamefont
  {Skomski}}, \bibinfo {author} {\bibfnamefont {P.}~\bibnamefont {Manchanda}},\
  and\ \bibinfo {author} {\bibfnamefont {A.}~\bibnamefont {Kashyap}},\ }\href
  {https://doi.org/10.1007/978-3-030-63101-7_3-1} {\bibfield  {journal}
  {\bibinfo  {journal} {Handbook of Magnetism and Magnetic Materials}\ ,\
  \bibinfo {pages} {1}} (\bibinfo {year} {2020})}\BibitemShut {NoStop}%
\bibitem [{\citenamefont {Kuz'min}\ and\ \citenamefont
  {Tishin}(2007)}]{kuzmin2007chapter}%
  \BibitemOpen
  \bibfield  {author} {\bibinfo {author} {\bibfnamefont {M.}~\bibnamefont
  {Kuz'min}}\ and\ \bibinfo {author} {\bibfnamefont {A.}~\bibnamefont
  {Tishin}},\ }\href {https://doi.org/10.1016/S1567-2719(07)17003-7} {\bibfield
   {journal} {\bibinfo  {journal} {Handbook of Magnetic Materials}\ }\textbf
  {\bibinfo {volume} {17}},\ \bibinfo {pages} {149} (\bibinfo {year}
  {2007})}\BibitemShut {NoStop}%
\bibitem [{\citenamefont {Radwa{\'n}ski}\ and\ \citenamefont
  {Franse}(1989)}]{radwanski1989magnetocrystalline}%
  \BibitemOpen
  \bibfield  {author} {\bibinfo {author} {\bibfnamefont {R.}~\bibnamefont
  {Radwa{\'n}ski}}\ and\ \bibinfo {author} {\bibfnamefont {J.}~\bibnamefont
  {Franse}},\ }\href {https://doi.org/10.1016/0921-4526(89)90066-5} {\bibfield
  {journal} {\bibinfo  {journal} {Physica B: Condensed Matter}\ }\textbf
  {\bibinfo {volume} {154}},\ \bibinfo {pages} {181} (\bibinfo {year}
  {1989})}\BibitemShut {NoStop}%
\bibitem [{\citenamefont {Stevens}(1952)}]{stevens1952matrix}%
  \BibitemOpen
  \bibfield  {author} {\bibinfo {author} {\bibfnamefont {K.}~\bibnamefont
  {Stevens}},\ }\href {https://doi.org/10.1088/0370-1298/65/3/308} {\bibfield
  {journal} {\bibinfo  {journal} {Proceedings of the Physical Society. Section
  A}\ }\textbf {\bibinfo {volume} {65}},\ \bibinfo {pages} {209} (\bibinfo
  {year} {1952})}\BibitemShut {NoStop}%
\bibitem [{\citenamefont {Patrick}\ \emph {et~al.}(2020)\citenamefont
  {Patrick}, \citenamefont {Marchant},\ and\ \citenamefont
  {Staunton}}]{Patrick2020}%
  \BibitemOpen
  \bibfield  {author} {\bibinfo {author} {\bibfnamefont {C.~E.}\ \bibnamefont
  {Patrick}}, \bibinfo {author} {\bibfnamefont {G.~A.}\ \bibnamefont
  {Marchant}},\ and\ \bibinfo {author} {\bibfnamefont {J.~B.}\ \bibnamefont
  {Staunton}},\ }\href {https://doi.org/10.1103/PhysRevApplied.14.014091}
  {\bibfield  {journal} {\bibinfo  {journal} {Phys. Rev. Applied}\ }\textbf
  {\bibinfo {volume} {14}},\ \bibinfo {pages} {014091} (\bibinfo {year}
  {2020})}\BibitemShut {NoStop}%
\bibitem [{\citenamefont {Yamada}\ \emph {et~al.}(1988)\citenamefont {Yamada},
  \citenamefont {Kato}, \citenamefont {Yamamoto},\ and\ \citenamefont
  {Nakagawa}}]{Yamada}%
  \BibitemOpen
  \bibfield  {author} {\bibinfo {author} {\bibfnamefont {M.}~\bibnamefont
  {Yamada}}, \bibinfo {author} {\bibfnamefont {H.}~\bibnamefont {Kato}},
  \bibinfo {author} {\bibfnamefont {H.}~\bibnamefont {Yamamoto}},\ and\
  \bibinfo {author} {\bibfnamefont {Y.}~\bibnamefont {Nakagawa}},\ }\href
  {https://doi.org/10.1103/PhysRevB.38.620} {\bibfield  {journal} {\bibinfo
  {journal} {Phys. Rev. B}\ }\textbf {\bibinfo {volume} {38}},\ \bibinfo
  {pages} {620} (\bibinfo {year} {1988})}\BibitemShut {NoStop}%
\bibitem [{\citenamefont {Gatteschi}\ and\ \citenamefont
  {Sessoli}(2003)}]{gatteschi2003quantum}%
  \BibitemOpen
  \bibfield  {author} {\bibinfo {author} {\bibfnamefont {D.}~\bibnamefont
  {Gatteschi}}\ and\ \bibinfo {author} {\bibfnamefont {R.}~\bibnamefont
  {Sessoli}},\ }\href {https://doi.org/10.1002/anie.200390099} {\bibfield
  {journal} {\bibinfo  {journal} {Angewandte Chemie International Edition}\
  }\textbf {\bibinfo {volume} {42}},\ \bibinfo {pages} {268} (\bibinfo {year}
  {2003})}\BibitemShut {NoStop}%
\bibitem [{\citenamefont {Schumacher}\ \emph {et~al.}(2014)\citenamefont
  {Schumacher}, \citenamefont {Huttmann}, \citenamefont
  {Petrovi\ifmmode~\acute{c}\else \'{c}\fi{}}, \citenamefont {Witt},
  \citenamefont {F\"orster}, \citenamefont {Vo-Van}, \citenamefont {Coraux},
  \citenamefont {Mart\'{\i}nez-Galera}, \citenamefont {Sessi}, \citenamefont
  {Vergara}, \citenamefont {R\"uckamp}, \citenamefont {Gr\"uninger},
  \citenamefont {Schleheck}, \citenamefont {Meyer~{zu Heringdorf}},
  \citenamefont {Ohresser}, \citenamefont {Kralj}, \citenamefont {Wehling},\
  and\ \citenamefont {Michely}}]{Eu_michely}%
  \BibitemOpen
  \bibfield  {author} {\bibinfo {author} {\bibfnamefont {S.}~\bibnamefont
  {Schumacher}}, \bibinfo {author} {\bibfnamefont {F.}~\bibnamefont
  {Huttmann}}, \bibinfo {author} {\bibfnamefont {M.}~\bibnamefont
  {Petrovi\ifmmode~\acute{c}\else \'{c}\fi{}}}, \bibinfo {author}
  {\bibfnamefont {C.}~\bibnamefont {Witt}}, \bibinfo {author} {\bibfnamefont
  {D.~F.}\ \bibnamefont {F\"orster}}, \bibinfo {author} {\bibfnamefont
  {C.}~\bibnamefont {Vo-Van}}, \bibinfo {author} {\bibfnamefont
  {J.}~\bibnamefont {Coraux}}, \bibinfo {author} {\bibfnamefont {A.~J.}\
  \bibnamefont {Mart\'{\i}nez-Galera}}, \bibinfo {author} {\bibfnamefont
  {V.}~\bibnamefont {Sessi}}, \bibinfo {author} {\bibfnamefont
  {I.}~\bibnamefont {Vergara}}, \bibinfo {author} {\bibfnamefont
  {R.}~\bibnamefont {R\"uckamp}}, \bibinfo {author} {\bibfnamefont
  {M.}~\bibnamefont {Gr\"uninger}}, \bibinfo {author} {\bibfnamefont
  {N.}~\bibnamefont {Schleheck}}, \bibinfo {author} {\bibfnamefont
  {F.}~\bibnamefont {Meyer~{zu Heringdorf}}}, \bibinfo {author} {\bibfnamefont
  {P.}~\bibnamefont {Ohresser}}, \bibinfo {author} {\bibfnamefont
  {M.}~\bibnamefont {Kralj}}, \bibinfo {author} {\bibfnamefont {T.~O.}\
  \bibnamefont {Wehling}},\ and\ \bibinfo {author} {\bibfnamefont
  {T.}~\bibnamefont {Michely}},\ }\href
  {https://doi.org/10.1103/PhysRevB.90.235437} {\bibfield  {journal} {\bibinfo
  {journal} {Phys. Rev. B}\ }\textbf {\bibinfo {volume} {90}},\ \bibinfo
  {pages} {235437} (\bibinfo {year} {2014})}\BibitemShut {NoStop}%
\bibitem [{\citenamefont {Abdelouahed}\ \emph {et~al.}(2007)\citenamefont
  {Abdelouahed}, \citenamefont {Baadji},\ and\ \citenamefont
  {Alouani}}]{Gd_orb}%
  \BibitemOpen
  \bibfield  {author} {\bibinfo {author} {\bibfnamefont {S.}~\bibnamefont
  {Abdelouahed}}, \bibinfo {author} {\bibfnamefont {N.}~\bibnamefont
  {Baadji}},\ and\ \bibinfo {author} {\bibfnamefont {M.}~\bibnamefont
  {Alouani}},\ }\href {https://doi.org/10.1103/PhysRevB.75.094428} {\bibfield
  {journal} {\bibinfo  {journal} {Phys. Rev. B}\ }\textbf {\bibinfo {volume}
  {75}},\ \bibinfo {pages} {094428} (\bibinfo {year} {2007})}\BibitemShut
  {NoStop}%
\bibitem [{\citenamefont {Jiang}\ and\ \citenamefont
  {Qin}(2015)}]{jiang2015prediction}%
  \BibitemOpen
  \bibfield  {author} {\bibinfo {author} {\bibfnamefont {S.}~\bibnamefont
  {Jiang}}\ and\ \bibinfo {author} {\bibfnamefont {S.}~\bibnamefont {Qin}},\
  }\href {https://doi.org/10.1039/C5QI00052A} {\bibfield  {journal} {\bibinfo
  {journal} {Inorganic Chemistry Frontiers}\ }\textbf {\bibinfo {volume} {2}},\
  \bibinfo {pages} {613} (\bibinfo {year} {2015})}\BibitemShut {NoStop}%
\bibitem [{\citenamefont {Sievers}(1982)}]{sievers1982asphericity}%
  \BibitemOpen
  \bibfield  {author} {\bibinfo {author} {\bibfnamefont {J.}~\bibnamefont
  {Sievers}},\ }\href {https://doi.org/https://doi.org/10.1007/BF01321865}
  {\bibfield  {journal} {\bibinfo  {journal} {Zeitschrift f{\"u}r Physik B
  Condensed Matter}\ }\textbf {\bibinfo {volume} {45}},\ \bibinfo {pages} {289}
  (\bibinfo {year} {1982})}\BibitemShut {NoStop}%
\bibitem [{\citenamefont {Kraus}\ \emph {et~al.}(2022)\citenamefont {Kraus},
  \citenamefont {Herman}, \citenamefont {Huttmann}, \citenamefont {Kramer},
  \citenamefont {Amsharov}, \citenamefont {Tsukamoto}, \citenamefont {Wende},
  \citenamefont {Atodiresei},\ and\ \citenamefont
  {Michely}}]{kraus2022selecting}%
  \BibitemOpen
  \bibfield  {author} {\bibinfo {author} {\bibfnamefont {S.}~\bibnamefont
  {Kraus}}, \bibinfo {author} {\bibfnamefont {A.}~\bibnamefont {Herman}},
  \bibinfo {author} {\bibfnamefont {F.}~\bibnamefont {Huttmann}}, \bibinfo
  {author} {\bibfnamefont {C.}~\bibnamefont {Kramer}}, \bibinfo {author}
  {\bibfnamefont {K.}~\bibnamefont {Amsharov}}, \bibinfo {author}
  {\bibfnamefont {S.}~\bibnamefont {Tsukamoto}}, \bibinfo {author}
  {\bibfnamefont {H.}~\bibnamefont {Wende}}, \bibinfo {author} {\bibfnamefont
  {N.}~\bibnamefont {Atodiresei}},\ and\ \bibinfo {author} {\bibfnamefont
  {T.}~\bibnamefont {Michely}},\ }\href {https://doi.org/10.1021/jacs.2c04359}
  {\bibfield  {journal} {\bibinfo  {journal} {Journal of the American Chemical
  Society}\ }\textbf {\bibinfo {volume} {144}},\ \bibinfo {pages} {11003}
  (\bibinfo {year} {2022})}\BibitemShut {NoStop}%
\bibitem [{\citenamefont {Huttmann}\ \emph {et~al.}(2015)\citenamefont
  {Huttmann}, \citenamefont {Mart{\'i}nez-Galera}, \citenamefont {Caciuc},
  \citenamefont {Atodiresei}, \citenamefont {Schumacher}, \citenamefont
  {Standop}, \citenamefont {Hamada}, \citenamefont {Wehling}, \citenamefont
  {Bl{\"u}gel},\ and\ \citenamefont {Michely}}]{huttmann2015tuning}%
  \BibitemOpen
  \bibfield  {author} {\bibinfo {author} {\bibfnamefont {F.}~\bibnamefont
  {Huttmann}}, \bibinfo {author} {\bibfnamefont {A.~J.}\ \bibnamefont
  {Mart{\'i}nez-Galera}}, \bibinfo {author} {\bibfnamefont {V.}~\bibnamefont
  {Caciuc}}, \bibinfo {author} {\bibfnamefont {N.}~\bibnamefont {Atodiresei}},
  \bibinfo {author} {\bibfnamefont {S.}~\bibnamefont {Schumacher}}, \bibinfo
  {author} {\bibfnamefont {S.}~\bibnamefont {Standop}}, \bibinfo {author}
  {\bibfnamefont {I.}~\bibnamefont {Hamada}}, \bibinfo {author} {\bibfnamefont
  {T.~O.}\ \bibnamefont {Wehling}}, \bibinfo {author} {\bibfnamefont
  {S.}~\bibnamefont {Bl{\"u}gel}},\ and\ \bibinfo {author} {\bibfnamefont
  {T.}~\bibnamefont {Michely}},\ }\href
  {https://doi.org/10.1103/PhysRevLett.115.236101} {\bibfield  {journal}
  {\bibinfo  {journal} {Physical review letters}\ }\textbf {\bibinfo {volume}
  {115}},\ \bibinfo {pages} {236101} (\bibinfo {year} {2015})}\BibitemShut
  {NoStop}%
\bibitem [{\citenamefont {Schumacher}\ \emph {et~al.}(2013)\citenamefont
  {Schumacher}, \citenamefont {Wehling}, \citenamefont {Lazic}, \citenamefont
  {Runte}, \citenamefont {F{\"o}rster}, \citenamefont {Busse}, \citenamefont
  {Petrovic}, \citenamefont {Kralj}, \citenamefont {Blu{\"u}gel}, \citenamefont
  {Atodiresei}, \citenamefont {Caciuc},\ and\ \citenamefont
  {Michely}}]{schumacher2013backside}%
  \BibitemOpen
  \bibfield  {author} {\bibinfo {author} {\bibfnamefont {S.}~\bibnamefont
  {Schumacher}}, \bibinfo {author} {\bibfnamefont {T.~O.}\ \bibnamefont
  {Wehling}}, \bibinfo {author} {\bibfnamefont {P.}~\bibnamefont {Lazic}},
  \bibinfo {author} {\bibfnamefont {S.}~\bibnamefont {Runte}}, \bibinfo
  {author} {\bibfnamefont {D.~F.}\ \bibnamefont {F{\"o}rster}}, \bibinfo
  {author} {\bibfnamefont {C.}~\bibnamefont {Busse}}, \bibinfo {author}
  {\bibfnamefont {M.}~\bibnamefont {Petrovic}}, \bibinfo {author}
  {\bibfnamefont {M.}~\bibnamefont {Kralj}}, \bibinfo {author} {\bibfnamefont
  {S.}~\bibnamefont {Blu{\"u}gel}}, \bibinfo {author} {\bibfnamefont
  {N.}~\bibnamefont {Atodiresei}}, \bibinfo {author} {\bibfnamefont
  {V.}~\bibnamefont {Caciuc}},\ and\ \bibinfo {author} {\bibfnamefont
  {T.}~\bibnamefont {Michely}},\ }\href {https://doi.org/10.1021/nl402797j}
  {\bibfield  {journal} {\bibinfo  {journal} {Nano letters}\ }\textbf {\bibinfo
  {volume} {13}},\ \bibinfo {pages} {5013} (\bibinfo {year}
  {2013})}\BibitemShut {NoStop}%
\bibitem [{\citenamefont {Cenker}\ \emph {et~al.}(2022)\citenamefont {Cenker},
  \citenamefont {Sivakumar}, \citenamefont {Xie}, \citenamefont {Miller},
  \citenamefont {Thijssen}, \citenamefont {Liu}, \citenamefont {Dismukes},
  \citenamefont {Fonseca}, \citenamefont {Anderson}, \citenamefont {Zhu},
  \citenamefont {Roy}, \citenamefont {Xiao}, \citenamefont {Chu}, \citenamefont
  {Cao},\ and\ \citenamefont {Xu}}]{cenker2022reversible}%
  \BibitemOpen
  \bibfield  {author} {\bibinfo {author} {\bibfnamefont {J.}~\bibnamefont
  {Cenker}}, \bibinfo {author} {\bibfnamefont {S.}~\bibnamefont {Sivakumar}},
  \bibinfo {author} {\bibfnamefont {K.}~\bibnamefont {Xie}}, \bibinfo {author}
  {\bibfnamefont {A.}~\bibnamefont {Miller}}, \bibinfo {author} {\bibfnamefont
  {P.}~\bibnamefont {Thijssen}}, \bibinfo {author} {\bibfnamefont
  {Z.}~\bibnamefont {Liu}}, \bibinfo {author} {\bibfnamefont {A.}~\bibnamefont
  {Dismukes}}, \bibinfo {author} {\bibfnamefont {J.}~\bibnamefont {Fonseca}},
  \bibinfo {author} {\bibfnamefont {E.}~\bibnamefont {Anderson}}, \bibinfo
  {author} {\bibfnamefont {X.}~\bibnamefont {Zhu}}, \bibinfo {author}
  {\bibfnamefont {X.}~\bibnamefont {Roy}}, \bibinfo {author} {\bibfnamefont
  {D.}~\bibnamefont {Xiao}}, \bibinfo {author} {\bibfnamefont {J.}~\bibnamefont
  {Chu}}, \bibinfo {author} {\bibfnamefont {T.}~\bibnamefont {Cao}},\ and\
  \bibinfo {author} {\bibfnamefont {X.}~\bibnamefont {Xu}},\ }\href
  {https://doi.org/10.1038/s41565-021-01052-6} {\bibfield  {journal} {\bibinfo
  {journal} {Nature Nanotechnology}\ }\textbf {\bibinfo {volume} {17}},\
  \bibinfo {pages} {256} (\bibinfo {year} {2022})}\BibitemShut {NoStop}%
\bibitem [{\citenamefont {Jiles}\ and\ \citenamefont
  {Lo}(2003)}]{jiles2003role}%
  \BibitemOpen
  \bibfield  {author} {\bibinfo {author} {\bibfnamefont {D.}~\bibnamefont
  {Jiles}}\ and\ \bibinfo {author} {\bibfnamefont {C.}~\bibnamefont {Lo}},\
  }\href {https://doi.org/10.1016/S0924-4247(03)00255-3} {\bibfield  {journal}
  {\bibinfo  {journal} {Sensors and Actuators A: Physical}\ }\textbf {\bibinfo
  {volume} {106}},\ \bibinfo {pages} {3} (\bibinfo {year} {2003})}\BibitemShut
  {NoStop}%
\bibitem [{\citenamefont {Hu}\ \emph {et~al.}(2020)\citenamefont {Hu},
  \citenamefont {Zhao}, \citenamefont {Krasheninnikov}, \citenamefont {Chen},\
  and\ \citenamefont {Sun}}]{hu2020enhanced}%
  \BibitemOpen
  \bibfield  {author} {\bibinfo {author} {\bibfnamefont {X.}~\bibnamefont
  {Hu}}, \bibinfo {author} {\bibfnamefont {X.}~\bibnamefont {Zhao},
  \bibfnamefont {Y.and~Shen}}, \bibinfo {author} {\bibfnamefont {A.~V.}\
  \bibnamefont {Krasheninnikov}}, \bibinfo {author} {\bibfnamefont
  {Z.}~\bibnamefont {Chen}},\ and\ \bibinfo {author} {\bibfnamefont
  {L.}~\bibnamefont {Sun}},\ }\href {https://doi.org/10.1021/acsami.0c05530}
  {\bibfield  {journal} {\bibinfo  {journal} {ACS applied materials \&
  interfaces}\ }\textbf {\bibinfo {volume} {12}},\ \bibinfo {pages} {26367}
  (\bibinfo {year} {2020})}\BibitemShut {NoStop}%
\bibitem [{\citenamefont {Morse}(1929)}]{Morse1}%
  \BibitemOpen
  \bibfield  {author} {\bibinfo {author} {\bibfnamefont {P.~M.}\ \bibnamefont
  {Morse}},\ }\href {https://doi.org/10.1103/PhysRev.34.57} {\bibfield
  {journal} {\bibinfo  {journal} {Phys. Rev.}\ }\textbf {\bibinfo {volume}
  {34}},\ \bibinfo {pages} {57} (\bibinfo {year} {1929})}\BibitemShut {NoStop}%
\bibitem [{\citenamefont {Minkin}(1999)}]{Morse2}%
  \BibitemOpen
  \bibfield  {author} {\bibinfo {author} {\bibfnamefont {V.~I.}\ \bibnamefont
  {Minkin}},\ }\href {https://doi.org/10.1351/pac199971101919} {\bibfield
  {journal} {\bibinfo  {journal} {Pure and Applied Chemistry}\ }\textbf
  {\bibinfo {volume} {71}},\ \bibinfo {pages} {1919} (\bibinfo {year}
  {1999})}\BibitemShut {NoStop}%
\end{thebibliography}%

\end{document}